# Propositions

accompanying the dissertation

**METRIC OPTIMIZATION AND MAINSTREAM BIAS MITIGATION IN RECOMMENDER SYSTEMS**

by

**Roger Zhe LI**

1. Bias roots in data, but its effect depends on the model.
   (**this thesis**)

2. Hyperparameters are abused because of sloppy mathematical modeling.
   (**this thesis**)

3. Equal treatment of users when training does not result in equal quality when recommending.
   (**this thesis**)

4. Investing in GPUs is less important than investing in data.

5. Deadlines are effective only when set by others.

6. Experience in industry makes better PhD candidates in technical sciences.

7. Globalization helps local diversity but ends up undermining global diversity.

8. PhD candidates will graduate much faster with proper psychological support.

9. Transferable skill courses provided by the graduate school are useless if not embedded into the academic context.

10. Publishing code next to papers does not help reproducibility as much as it reveals the mismatch between what research is done and what is reported.

These propositions are regarded as opposable and defendable, and have been approved as such by the promotors prof. dr. A. Hanjalic and dr. J. Urbano.

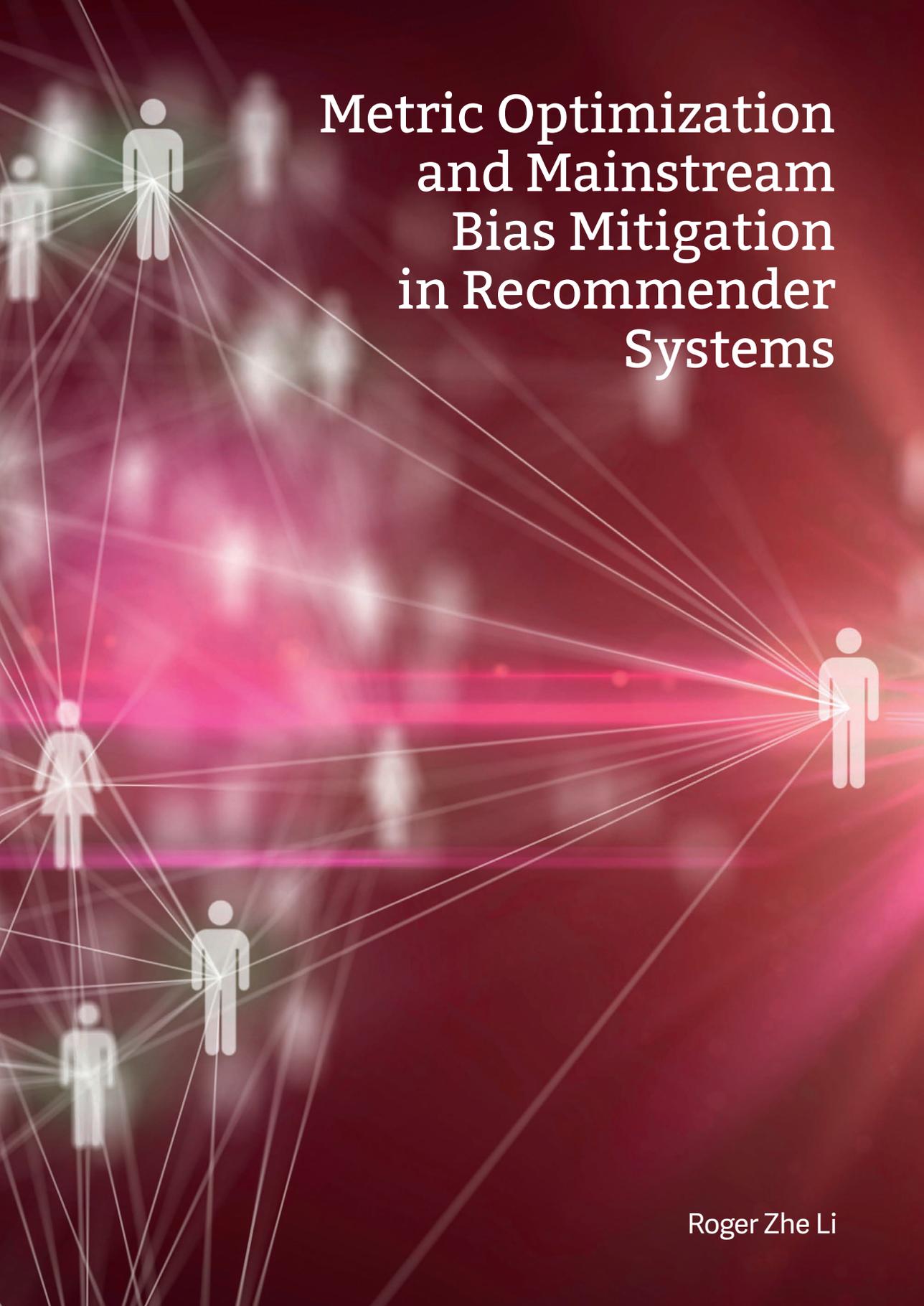

# Metric Optimization and Mainstream Bias Mitigation in Recommender Systems

Roger Zhe Li

# METRIC OPTIMIZATION AND MAINSTREAM BIAS MITIGATION IN RECOMMENDER SYSTEMS

# Metric Optimization and Mainstream Bias Mitigation in Recommender Systems

**Proefschrift**

door

## Zhe Lı

Master of Sciences in Information and Communication Engineering,
Tianjin University, Tianjin, Volksrepubliek China,
geboren te Taiyuan, Volksrepubliek China.

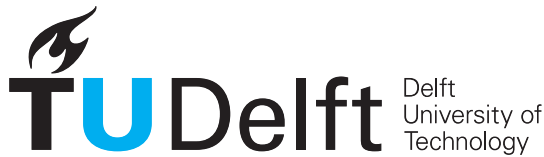

*To be yourself in a world that is constantly trying to make you something else is the greatest accomplishment.*

Ralph Waldo Emerson

# CONTENTS

# 1

# Introduction

## 1.1. RECOMMENDER SYSTEMS

With the increasing number of information channels available to us nowadays, one can easily realize that more is not always better. The problem of information overload [50] has never been bigger, and with that also the challenge of filtering out the information that is useful to us [111, 25]. The same holds for numerous online transactions platforms, where we are exposed to many options to choose from (e.g., booking a hotel), or many items to buy. The technology that proved to be helpful in this respect is the one underlying the *recommender systems* [134, 147]. Such systems have been successfully applied in numerous application scenarios including e-commerce [136, 72], entertainment [45, 16], news communication [141], social media [121], education [58], healthcare [124], and Internet of Things (IoT) [39]. Their core function is to predict the need of a customer (in the remainder of this thesis referred to as *user*) based on a learned preference model, and to provide a recommendation service that best satisfies that need. Since this service can be targeted towards recommending various things, from (trans)actions, via information, to things to buy, in the remainder of this thesis, we will refer to the targets of this service as *items*.

Recommender systems by nature involve multiple stakeholders [3]. Next to the primary goal to provide service that satisfies their users, the commercial and strategic interests of online product sellers and system (platform) providers (e.g., Amazon, Booking.com) need to be satisfied as well. For example, in a system recommending information and services in the tourism domain [17, 92], users may seek to book a hotel that matches their interests, and that is both affordable and convenient, while hotels seek to maximize their profit margin. At the same time, the platform aims at attracting more hotels as well as satisfying the exposure needs of those who paid the commission [55]. An ideal recommender system is thus expected to find an operating strategy that is good for all parties involved, with the challenge that the interests of those parties are not necessarily aligned with each other. Since the users' satisfaction can be seen as the precondition of a successfully operating recommender system, is has been the main focus in the development of these systems.

Users of recommender systems typically have heterogeneous demands, which makes user satisfaction an ensemble of multiple aspects. Starting with the demand that the recommended information should be *relevant* to the users in the first place (i.e., corresponding to the user's need in general), other aspects, like diversity [24, 117, 151], novelty [145], and serendipity [54] may also play an important role, as well as self-actualization [59], which helps the users get new insights into their preferences and develop new preferential patterns for the future.

## 1.2. ACCURACY-ORIENTED RECOMMENDATION

Since there is no good recommendation without relevant items, one can argue that the prerequisite of user satisfaction is to get relevant items pushed to the top of the list of recommended items (also referred to as *recommendation list*). The ability of a recommender system to achieve this is also referred to as recommendation *accuracy*. The other aspects mentioned above, that we refer to as *beyond-accuracy* demands, can only be meaningfully addressed if a sufficient relevance base is secured. For example, diversify-ing among non-relevant items does not make a user more satisfied with the service, and self-actualization can only be meaningfully deployed if the users are confident about the system's ability to understand their initial needs and the evolution of these needs.

In view of the above, the major scientific and development effort in the field of recommender systems has been invested in maximizing the recommendation accuracy. This translates onto the development of machine learning models that exploit previous interaction history of a user (recommendation via *content-based filtering*) and other users (recommendation via *collaborative filtering*) with items, as well as any other available side information on users and items to either predict the ratings of a user to new items, or predict how a user would rank new relevant items. Developing a recommendation model is based on the following main considerations:

- **Collecting abundant data to learn from**. In recommender systems based on collaborative filtering (our focus in this thesis, further denoted by CF), previous user-item interactions are captured by a user-item matrix (UIM). Such matrix can contain explicit ratings, or signals implicitly representing user preferences (e.g., watching or downloading an item, time spent on analyzing an item). The more the users are active in such interactions, the more information is captured in the UIM. In addition to UIM, there are, in general, other information resources to deploy, such as the information on the users (e.g., demographics) and items (e.g. content features) themselves, and about different use contexts (e.g., user preferences varying per season), for more comprehensive preference modeling and better recommendation-model training [30, 114].

- **Defining the recommendation-learning model**. Collaborative Filtering, which is built upon the assumption that users sharing similar preferences in the past will keep having similar taste in the future, is the most successful and widely-used approach to build recommender systems. Depending on the type of the available input data, learning to recommend in the CF case relies in general on three categories of learning models: Matrix Factorization [60] (based on UIM only), Factorization Machines [99] (based on UIM accompanied by contextual information on users, items and their interactions), and deep neural networks (DNNs) (abundant information about users, items and their interactions). In particular, compared to the first two categories, DNNs are also capable of capturing higher-order user-user, user-item and item-item relations, resulting in more advanced learning models [47, 48].

- **Finding a criterion to optimize for**. Machine learning algorithms are generally developed to optimize a criterion defined as a function, that is, to maximize a reward or minimize a cost function. Since the choice and quality of the criterion plays a crucial role in achieving the best possible recommendation accuracy, its definition becomes a key issue in the development of recommender systems. Previous work adopts a number of accuracy-oriented criteria focusing on different aspects. In the case of *rating prediction*, the optimization criteria typically deployed are Mean Absolute Error (MAE) and Root Mean Square Error (RMSE), while *ranking prediction* typically deploys Reciprocal Rank (*RR*) [33], normalized Discounted Cumulative Gain (*nDCG*) [51] and Average Precision (*AP*) [81].

**1**

Within the above three items, finding a good criterion to optimize for may be the most tricky one in the attempt to provide a recommendation service leading to high satisfaction of as many users as possible. While a choice of a widely applied, popular optimization criterion could, on the first sight, appear as the best educated guess, this criterion may not tailor to the needs of all users. They may focus on different aspects and perceive the quality of the produced recommendation list differently. Research on the choice of the optimization criterion has received extensive attention in the domain of information retrieval, especially regarding direct optimization for a ranking-based evaluation metric. In the recommender systems community, Shi et al. [114, 113] and Liang et al. [71] showed that the evaluation criteria could be directly selected to optimize the recommendation model. On the other hand, both empirical study [135] and theoretical derivation based on the maximum entropy framework [10, 142] show that optimizing for some criteria would lead to a broader effectiveness of a retrieval system compared to other criteria. This may indicate that, in order to maximize the score on one evaluation criterion, it may not be necessary to also optimize for it. Instead, using a more generic optimization criterion may lead to even better performance, independent of which evaluation criterion is deployed. Inspired by this knowledge gap, the first focus of this thesis is to dive deep into this topic, check the effectiveness of optimizing a recommender system in view of different criteria, and find an answer to the first main research question underlying this thesis:

*RQ1: In order to achieve the best possible accuracy for a broad population of users, should we optimize a recommender system for the criterion we would evaluate it on?*

## **1.3. BIAS IN RECOMMENDATION**

It is not always possible to effectively address the challenge of maximizing the satisfaction of a broad user population by only focusing on selecting the appropriate optimization criterion. One of the main factors that typically needs to also be taken into account is the *bias* [27] in recommendations. While several categories of bias have been discussed in the literature [12], we focus in this thesis on the bias caused by imbalance in the user-related input data (further referred to as the user base), or more specifically, by the underrepresentation of some users or user groups in the data. This bias can be related to different factors, such as demographics (age [38], gender [69, 26], race [134]), and can lead to different sorts of effects on recommendation. For example, the imbalance in the user base in terms of gender can lead to serious consequences in some domains in which recommender systems are being deployed to automate processes, like hiring. Gender-based discrimination in job recruitment has been reported [8] as a consequence of correlating suitability for certain professions to gender when training the recommender system.

In this thesis, we focus on one typical scenario where imbalance in user base is caused by different taste and user-item interaction patterns across users. Intuitively, users who have different tastes from the majority are likely to interact with items that are less frequently purchased. Together with the users who are in general less active on the platform, these *non-mainstream users* generate less information for preference modeling than the majority or *mainstream users*. It can be said that, through the dominance of the mainstream users in steering the learning of a recommender system, non-mainstream users are more difficult to recommend to, and therefore suffer from the *mainstream bias* [15, 150]. Consequently, these users could be discouraged from using the systems in the first place [65]. This could make customer retention challenging for online businesses. For other use cases however, such as information and news recommendation, we foresee even more serious consequences. Recommender systems may become less inclusive with respect to non-mainstream opinions and views (e.g., political) and in this way contribute to undesired long-term effects, like intellectual segregation and societal polarization. In addition, the long-term existence of mainstream bias will result in a continuous improvement of the performance for the mainstream group as well as a continuous decrease of the performance for the rest [76], amplifying the accuracy gap between them even further.

From the technical perspective, the bias due to data imbalance emerges from how recommender systems are typically trained, namely by trying to minimize the recommendation error, or – as an equivalent – to maximize the accuracy averaged over all users. This averaging biases the learning of the recommendation model towards users who are overrepresented in the input data because their information is dominant in steering the learning process. Consequently, the minority group may become (partly) neglected. In other words, a good overall average accuracy score might be achieved even if the scores on the minority user group are bad, as long as those obtained for the users from the majority group are excellent. Since imbalance in input data is realistic in a practical use case, relying on the optimization of average accuracy is insufficient to develop effective recommender systems [12]. As an alternative, we need to care more about maximizing the accuracy of **every** user instead of their average. This idea of looking into the accuracy on the individual level when training a recommender system is closely related to the concept of risk-sensitive optimization [130] in information retrieval.

Despite being the subject of research over the past years, debiasing recommender systems remains a challenging research topic [27]. In this thesis, we give our contribution to this effort by searching for an answer to the following second main research question:

*RQ2: How to mitigate the mainstream bias in recommender systems, so that different users or user groups share a more balanced recommendation accuracy?*

## 1.4. CONTRIBUTIONS OF THIS THESIS

In the following, we explain how the core research questions stated above are addressed in different chapters of this thesis.

**Chapter** 2 focuses on the RQ1, namely the impact of choosing different optimization targets on the average recommendation accuracy. We present an extensive experimental study conducted on different datasets in both pairwise and listwise learning-to-rank (LTR) scenarios, to compare the relative merit of popular IR metrics used in the recommendation context, namely *RR*, *AP* and *nDCG*, with Rank-Biased Precision (*RBP*) [87] when used for optimization and assessment of recommender systems in various combinations. For the first three, we follow the practice of loss function formulation available in literature. For the fourth one, we propose novel loss functions inspired by *RBP* for both the pairwise and listwise scenario. Our results confirm that the best performance is indeed not necessarily achieved when optimizing the same metric being used for evalu-

**1**

ation. In fact, we find that *RBP*-inspired losses perform at least as well as other metrics in a consistent way, and offer clear benefits in several cases. Interesting to see is that *RBP*-inspired losses, while improving the recommendation performance for all users, may lead to an individual performance gain that is correlated with the activity level of a user in interacting with items. The more active the users, the more they benefit. Overall, our results challenge the assumption behind the current research practice of optimizing and evaluating the same metric, and point to *RBP*-based optimization instead as a promising alternative when learning to rank in the recommendation context.

Upon answering the RQ1 related to average recommendation accuracy, we focus in chapters 3 and 4 on the RQ2 related to the accuracy at the individual user level and propose ideas for mitigating the mainstream bias. In **Chapter 3**, we propose NAECF, a conceptually simple but effective idea to introduce extra information to learn from and to maintain the intrinsic features of all users in the recommendation model. The idea consists of adding an autoencoder (AE) layer when learning user and item representations with text-based Convolutional Neural Networks. The AEs, one for the users and one for the items, serve as adversaries to the process of minimizing the rating prediction error when learning how to recommend. They enforce that the unique properties of all users and items are sufficiently well incorporated and preserved in the learned representations. These representations, extracted as the bottlenecks of the corresponding AEs, are expected to be less biased towards mainstream users, and to provide more balanced recommendation utility across all users. Our experimental results confirm that non-mainstream users tend to receive better recommendations under our proposed model, while the recommendation quality for mainstream users is maintained. Our results emphasize the importance of deploying extensive content-based features, such as online reviews, in order to better represent users and items and thus maximize the de-biasing effect.

In **Chapter 4**, we explore a different avenue to mitigate the mainstream bias solely based on the user-item interactions under the learning-to-rank paradigm. Without external resources, we first investigate whether it is possible to make the importance of users explicitly related to their mainstreamness during training, and propose a simple user-weighting approach incorporated into the training process by taking the cost of potential recommendation errors into account. In this way, mainstreamness determines the relative importance of a user, and the model is pushed to focus more on non-mainstream users and improve their recommendation accuracy. We further discuss the way to define mainstreamness, and find that compared to other measures, recommendation accuracy could be a more direct and powerful proxy to quantify mainstreamness. Finally, since our work is built upon individual users rather than looking into the whole user base, data needs for such kind of research might be different. We therefore discuss how to get reliable results on evaluating recommendation quality from the data perspective, and provide suggestions regarding the minimum number of interactions per user when partitioning the datasets.

**Chapter 5** summarizes this thesis, draws conclusions, and points out the future research directions rooted in this thesis.

## 1.5. HOW TO READ THIS THESIS

The main scientific contribution of this thesis is presented in the technical chapters 2, 3 and 4. Each of these chapters is connected to one publication, which is referenced at the beginning of the chapter. In this book, we retain the original form of the publications, possibly with some minor modifications. Each technical chapter represents an independent work that can be read without necessarily reading previous chapters. As a consequence, there might be overlapping insights and background information in the introductory and related work sections of the technical chapters and the notation and terminology might vary slightly across them.

## 1.6. LIST OF PUBLICATIONS

The papers published in the course of the research towards this thesis are listed below. For the papers that are directly connected to this dissertation, the references to the corresponding technical chapters are added in parentheses.

- Roger Zhe Li, Julián Urbano, and Alan Hanjalic. "New Insights into Metric Optimization for Ranking-based Recommendation". In: *SIGIR*. ACM, 2021, pp. 932–941 (**Chapter 2**)

- Roger Zhe Li, Julián Urbano, and Alan Hanjalic. "Leave No User Behind: Towards Improving the Utility of Recommender Systems for Non-mainstream Users". In: *WSDM*. ACM, 2021, pp. 103–111 (**Chapter 3**)

- Roger Zhe Li, Julián Urbano, and Alan Hanjalic. "Mitigating Mainstream Bias in Recommendation via Cost-sensitive Learning". In: *ICTIR*. ACM, 2023, pp. 135–142 (**Chapter 4**)

- Zhe Li. "Towards the next generation of multi-criteria recommender systems". In: *RecSys*. ACM, 2018, pp. 553–557

# 2

# NEW INSIGHTS INTO METRIC OPTIMIZATION FOR RANKING-BASED RECOMMENDATION

## 2.1. INTRODUCTION

Offline evaluation of ranking-based recommender systems generally relies on effectiveness metrics from Information Retrieval (IR). These metrics quantify the quality of a ranked list of items in terms of their relevance [126] and according to the particular evaluation criteria they capture. Therefore, it has been seen as an intuitive and logical choice to learn a ranking model for a recommender system by directly optimizing the metric used for evaluation [73]. Different ranking approaches have been designed along this line, aiming at achieving a better recommendation performance [112, 114, 113, 71].

Previous research shows, however, that optimizing the metric used for evaluation is not necessarily the best approach. Different IR metrics reflect different aspects of retrieval performance [6, 53, 105, 85], and do so to different extents. Although no metric covers all evaluation criteria, there is evidence that some metrics are more informative than others [10, 142, 9]. This may enable them to, when used for optimization, achieve the best performance in view of a given evaluation metric other than itself, or even in view of multiple target evaluation criteria. Empirical results show that the advantage in informativeness can indeed be transferred into higher effectiveness when these metrics are used for optimization. Results on web search [35] and text retrieval [142] show that, when targeting at less informative metrics such as $P@10$ and $RR$, optimizing more informative metrics, like $AP$ or $nDCG$, can perform even better than optimizing the less informative metrics themselves.

The discussion above points to the possibility to achieve better ranking performance according to the evaluation criterion captured by metric $X$ if we do not optimize for $X$, but for a more informative metric $Y$ instead. This, however, is more likely to be successful when the metrics $X$ an $Y$ are highly correlated. From the perspective of ranking effectiveness, correlation would typically arise between metrics that share characteristics, such as top-weightedness and localization. According to Moffat [85], correlation in view of such properties could tell something about the ability of the metrics to reveal the same aspects of the system behavior, as opposed to non-correlated metrics that reveal different aspects of the system behavior. Previous research on text retrieval and web search has shown high level of correlation among many metrics, such as between $nDCG$ and $AP$, $RR$ and $AP$, or $P@k$ and $AP$ [135, 43, 11]. When targeting at maximizing one specific evaluation metric, one should therefore search among informative metrics correlated to the evaluation target to find the most effective optimization objective.

Motivated by the above, in this chapter we revisit the choice of the metric to optimize in a ranking-based recommendation scenario. Our goal is to provide more insights into this optimization scenario, broadening the possibilities to benefit from such a metric choice, compared to what is commonly reported in the literature.

Although several popular metrics like $RR$, $nDCG$ and $AP$ have been applied to the task, $RBP$ (Rank-Biased Precision) [87], another important effectiveness metric widely used in traditional IR tasks, has hardly been applied for training and/or testing recommendation models. Nonetheless, $RBP$ is an informative [85, 88, 7] and flexible metric that incorporates a simple user model through a *persistence* parameter. Since varying the value of the persistence parameter makes $RBP$ correlated with different groups of metrics [105, 86], $RBP$ has the potential to optimize for different metrics that reflect different evaluation criteria [87, 43].

Following the spirit of existing metric optimization processes, we propose novel objective functions inspired by *RBP* for both the pairwise and listwise learning-to-rank (LTR) paradigms. In this way, we enable *RBP* to join the other metrics and serve as both the learning objective and evaluation target for our investigation. Specifically for the listwise case, we will show that minimizing the proposed *RBP*-based objective function provides an elegant instrument to directly optimize for the ranking positions of the relevant items. Furthermore, the proposed *RBP*-based listwise loss function is independent of the persistence parameter *p*, which makes it possible to conduct *RBP*-based optimization using a single unified framework, regardless of the target persistence to evaluate.

Empirical results obtained on four real-world datasets point to the following main insights:

1. The assumption behind the practice to optimize and evaluate ranking-based recommender systems using the same metric does not necessarily lead to the best performance.

2. *RBP*-inspired losses perform at least as well as other metrics in a consistent way, and offer clear benefits in several cases. This makes *RBP* a promising alternative when learning to rank in the recommendation context.

3. *RBP*-based listwise optimization leads —relative to other metrics— to a significantly better ranking performance for active users with more of the relevant interactions, compared to less active users. However, this performance bias does *not* come at the cost of reducing utility for inactive users. On the contrary, the ranking performance improves for *all* users, only to different degrees.

The remainder of this chapter is organized as follows. In Section 2.2, we position our contribution in the context of the related previous work. Section 2.3 describes the technical details of the models we use for a direct optimization of IR metrics. Section 2.4 introduces the design and protocols of our experiments, the results of which we present and discuss in Section 2.5. Section 2.6 concludes the chapter with pointers to future work.

## 2.2. RELATED WORK

Direct optimization of IR metrics is a logical way of building ranking-based recommenders. Despite the fact that almost any IR metric can be transformed into an objective function for optimization, the choice of metric to optimize for maximizing the ranking effectiveness remains non-trivial. Intuitively, choosing to optimize more informative metrics helps with achieving higher ranking effectiveness. Several studies [10, 142, 9], inspired by Shannon [110] and Jaynes [52], have assessed the informativeness of IR metrics by how well they constrain a maximum entropy distribution over the relevance of ranked items, in the sense that such distribution accurately estimates the precision-recall curve or other metric scores for the same items. The results indicate that more complex metrics, such as *nDCG* and *AP*, are more informative than simpler metrics [135].

One way of achieving optimal ranking with respect to a given metric is to deploy a *pairwise* LTR approach. The pairwise LTR paradigm considers relevant-irrelevant (positive-negative) item pairs, and aims at maximizing the change in the considered IR metric caused by a ranking position swap. This idea lays the foundation for a batch of models, including LambdaRank [19], LambdaMart [21, 49] and LambdaFM [144]. In particular, LambdaRank is widely used as the underlying model in studies comparing the optimization of different metrics. LambdaRank-based results in [35, 142] show that optimizing for informative metrics can lead to good performance, not only when evaluating with the same metric, but also with others. This insight invites to revisit pairwise learning recommender systems by experimenting with other metrics to optimize, even if they are not the evaluation target. Similarly, LambdaFM [144] was deployed to assess the effectiveness with respect to three metrics, namely *nDCG*, *RR* and *AUC*, by optimizing for *nDCG*. Optimal performance was achieved with respect to *nDCG* and *RR*.

Another way of achieving optimal ranking with respect to a given metric is to deploy a *listwise* LTR approach. This approach looks at the entire ranked list for optimization, and therefore better resembles the concept of direct metric optimization than the (indirect) pairwise LTR approach. Although straightforward and close in nature to LTR, listwise methods have to deal with loss functions containing integer ranking positions, which causes non-smoothness and therefore non-differentiability. A common way to deal with this problem is to approximate the indicator function by a differentiable alternative. CofiRank [137] was one of the first works addressing this issue by choosing to minimize the $(1 - nDCG)$ loss with a structured estimation. Another popular method is to use a smooth function, such as a sigmoid or ReLU [91], to approximate the non-smooth indicator function. This method has been widely applied for optimizing *DCG* [71], *AP* [114] and *RR* [113]. Rather than optimizing the whole list and taking items at the bottom into account, Liang et al. [71] proposed Top-N-Rank, which focuses on the top ranked items and uses a listwise loss with a cutoff to directly optimize for $DCG@k$.

Despite this rich track record of attempts to learn a ranking by metric optimization, still insufficient is known about what metric to optimize for in order to obtain the best performance according to some evaluation metric. Moreover, we believe that the scope of the metrics to consider could further be expanded to broaden the possibilities for improving ranking effectiveness beyond what has been tried so far. In this work, to conduct our experimental assessment regarding ranking effectiveness for recommendation, we follow both the pairwise and listwise LTR approaches, and consider different IR metrics to optimize and assess ranking performance. Specifically, we add the *RBP* metric as a promising candidate to the set of typically deployed *RR*, *AP* and *nDCG*.

## 2.3. METHODS

In this section we describe the design choices and methodology behind our experimental approach to acquire new insights into the issues related to generating recommendations through optimizing IR metrics. We start by introducing our underlying recommendation model with the notation and terminology used throughout the chapter. Then, we describe the four IR metrics we choose to optimize. Finally, we define the corresponding objective functions we deploy for optimization in both the pairwise and listwise cases. In this way, we put special emphasis on the definitions of objective functions for *RBP* that we introduce in this chapter.

### 2.3.1. RECOMMENDATION MODEL

Recommender systems are meant to recommend "items" (in a general meaning of the term) to users according to their preferences. For a system with $M$ users and $N$ items, ground-truth user-item interaction data can be represented by a matrix $Y$ with dimensions $M \times N$. We consider a binary relevance scenario in this work, which implies that elements in $Y$ are either $y_{ui} = 1$, indicating a positive interaction (preference) between a user $u$ and an item $i$, or $y_{ui} = 0$, indicating either a negative interaction (e.g. a 'dislike'), or no interaction between $u$ and $i$. We refer to items with the positive (negative) interaction as the positive (negative) items. We assume that an arbitrary user $u$ generated $m_u$ positive interactions across all items.

Following the practice from other ranking-based recommendation approaches that target direct metric optimization [113, 71, 63], in this chapter we choose Matrix Factorization (MF) [101] as the recommendation model. Although collaborative filtering can be achieved via more advanced methods such as Neural Collaborative Filtering [47], Collaborative Variational Autoencoders [68] and Graph Neural Networks [48, 28], we still choose the base Matrix Factorization model because our aim in this chapter is to study the relative merits of metrics. A more comprehensive experiment to assess generalizability with other models is left for future work.

The users and items are thus represented by latent factor matrices $U^{M \times D}$ and $V^{N \times D}$, respectively, where $D$ is the number of latent factors. Using the latent vectors of users and items, a recommendation model can predict the relevance of items for each user, and store the scores in the matrix $F^{M \times N}$, with the element $f_{ui}$ representing the predicted relevance of item $i$ to user $u$. The ranking position $R_{ui}$ corresponding to the relevance score $f_{ui}$, is an integer ranging from 1 to $N$, calculated from a pairwise comparison between the predicted relevance score for item $i$ and all other items:

$$R_{ui} = 1 + \sum_{j=1 \setminus i}^{N} \mathbb{I}(f_{uj} > f_{ui}), \qquad (2.1)$$

where $\mathbb{I}(\cdot)$ denotes the indicator function.

### 2.3.2. METRICS

As introduced before, we consider four metrics to optimize when training the ranking mechanism of a recommender system: *RR*, *AP*, *nDCG* and *RBP*. These metrics, assessing the recommendation performance for user $u$, can be formulated as follows:

$$\begin{aligned} nDCG(u) &= \frac{DCG(u)}{iDCG(u)} \\ &= \frac{\sum_{i=1}^{N} (2^{y_{ui}} - 1) / \log_2(R_{ui} + 1)}{\sum_{i=1}^{m_u} 1 / \log_2(i + 1)}, \end{aligned} \qquad (2.2)$$

$$AP(u) = \frac{1}{m_u} \sum_{i=1}^{N} \frac{y_{ui}}{R_{ui}} \sum_{j=1}^{N} y_{uj} \mathbb{I}(R_{uj} \le R_{ui}), \qquad (2.3)$$

$$RR(u) = \sum_{i=1}^{N} \frac{y_{ui}}{R_{ui}} \prod_{j=1}^{N} \left(1 - y_{uj} \mathbb{I}(R_{uj} < R_{ui})\right), \qquad (2.4)$$

$$RBP(u;p) = (1-p) \sum_{i=1}^{N} y_{ui} p^{R_{ui}-1} \ . \tag{2.5}$$

According to the *RBP* formulation originally proposed by Moffat and Zobel [87], $p$ is a constant parameter ranging from 0 to 1, indicating the degree of persistence of a user. A high persistence models a user that is willing to explore items deep down the ranked list. The theoretical upper limit of *RBP* is 1 when *N* is infinite, which means 1 is never reached in practice [105]. To align the range of *RBP* with other metrics used in the chapter and in this way make it more comparable, we choose to optimize *nRBP* instead, which normalizes the bare *RBP* by the maximum obtainable with *m* positive items:

$$\begin{aligned} nRBP(u;p) &= \frac{RBP(u;p)}{iRBP(u;p)} \\ &= \frac{\sum_{i=1}^{N} y_{ui} p^{R_{ui}-1}}{\sum_{i=1}^{m_u} p^{i-1}} = Z(p, m_u) RBP(u;p) \ , \end{aligned} \tag{2.6}$$

where $Z(p, m_u) = 1/(1 - p^{m_u})$, serving as a normalization factor.

### 2.3.3. PAIRWISE METRIC OPTIMIZATION

Following the same rationale as in Section 2.3.1, we choose LambdaRank [19], the base Lambda gradient ranking model, as the pairwise LTR approach. We do not consider the more complex LambdaMART [21] and LambdaFM [144] to avoid the effect of other factors such as model ensemble and dynamic negative sampling strategy. Derived from RankNet [20], LambdaRank aims at obtaining smooth gradients for optimization by calculating the performance gain from swapping the position of documents $i$ and $j$ in a ranked list. For the $\lambda$-optimization of *RR*, *AP* and *nDCG*, we follow the existing approaches proposed by Donmez, Svore, and Burges [35]. To the best of our knowledge, LambdaRank using *RBP* has not been formally proposed yet, so we define the $\lambda$-optimization of *nRBP* following the same principles.

The cost for $\lambda$-optimizing an item pair $(i, j)$ for user $u$ is

$$C_{uij} = -S_{uij} o_{uij} + \ln\left(1 + e^{S_{uij} o_{uij}}\right) \ , \tag{2.7}$$

where $S_{uij}$ equals +1 or −1 depending on whether the ground truth label $y_{ui}$ is larger than that of $y_{uj}$, and the term $o_{uij} \equiv f_{ui} - f_{uj}$ represents the difference of the predicted relevance scores. The derivative of the cost with respect to $o_{uij}$ can be formulated as

$$\frac{\delta C_{uij}}{\delta o_{uij}} = -S_{uij} + \frac{S_{uij} e^{S_{uij} o_{uij}}}{1 + e^{S_{uij} o_{uij}}} = -\frac{S_{uij}}{1 + e^{S_{uij} o_{uij}}} \ . \tag{2.8}$$

In order to reward the positive gains and punish the negative, the $\lambda$-gradient for *nRBP* can thus be written as

$$\lambda_{uij} = S_{uij} \left| \Delta nRBP(R_{ui}, R_{uj}; p) \frac{\delta C_{uij}}{\delta o_{uij}} \right| \ , \tag{2.9}$$

where $R_{ui}$ and $R_{uj}$ are the ranking positions of the item pair, calculated as in Eq. (2.1), and $\Delta nRBP(R_{ui}, R_{uj}; p)$ is the difference between the corresponding nRBP values. This leads to the $RBP$-based $\lambda$-gradient formulated as

$$\lambda_{uij} = S_{uij} \left| \frac{Z(p, m_u)(1-p)\left(y_{ui} p^{R_{ui}-1} - y_{uj} p^{R_{uj}-1}\right)}{1 + e^{S_{uij} o_{uij}}} \right|. \tag{2.10}$$

### 2.3.4. LISTWISE METRIC OPTIMIZATION

Pairwise methods, such as LambdaRank, can easily avoid the issue of non-smoothness of the optimized ranking metrics. However, and despite their success, eluding this issue in the listwise approach remains an open problem [35]. Successfully addressing this challenge is important because, in that way, the optimization process becomes more intuitive, straightforward and natural [74]. As mentioned in Section 2.2, methods have already been proposed for listwise optimization of $RR$, $AP$ and $nDCG$. The underlying principle is to approximate the non-differentiable indicator function with a smooth alternative. This has been done by deploying either a sigmoid function [114, 113] or the Rectified Linear Unit (ReLU) [91] as proposed in [71]. To make a consistent comparison, in this chapter we use a sigmoid for all metrics.

Apart from the difference in the choice of the smoothing function, compared to TFMAP [114] and CLiMF [113], Top-N-Rank [71] is also distinctive for the way the approximated ranking position $\tilde{R}_{ui}$ is modeled. TFMAP and CLiMF model $\tilde{R}_{ui}$ using only the predicted score $f_{ui}$. In contrast, Top-N-Rank infers the predicted ranking position by looking at the whole recommendation list. It follows the idea from Eq. (2.1) to get the ranking position by pairwise comparison, which is closer in nature to sorting.

Taking all the above into account, and following the same rationale as in Sections 2.3.1 and 2.3.3, we do not contemplate more complex techniques such as Boosting [127, 40] or multi-agent learning [152], so that metrics are compared on a base recommendation model derived from Top-N-Rank. Specifically, we replace ReLU by a sigmoid function and approximate the ranking position as

$$\tilde{R}_{ui} = 1 + \sum_{j=1 \setminus i}^{N} \sigma(f_{uj} - f_{ui}), \tag{2.11}$$

where $\sigma(x) = 1/(1 + e^{-x})$. Accordingly, the approximation of the indicator functions in Eqs. (2.3) and (2.4) can be formulated as

$$\mathbb{I}(R_{uj} < R_{ui}) = \mathbb{I}(f_{uj} > f_{ui}) \approx \sigma(f_{uj} - f_{ui}), \tag{2.12}$$

$$\sum_{j=1}^{N} \mathbb{I}(R_{uj} \leq R_{ui}) = 1 + \sum_{j=1}^{N} \mathbb{I}(R_{uj} < R_{ui}) \approx 1 + \sum_{j=1 \setminus i}^{N} \sigma(f_{uj} - f_{ui}). \tag{2.13}$$

Top-N-Rank, however, optimizes metrics with a cutoff, so it does not use information from all items. As indicated by Donmez, Svore, and Burges [35], optimizing $nDCG$ on the whole item list can lead to significantly better $nDCG@10$ performance than directly optimizing $nDCG@10$. Consequently, we choose to eliminate the cutoff. Since the target is to maximize the approximated IR metrics, we can consider their additive inverse as the loss functions for optimization. Based on the above, the *RR*, *AP* and *nDCG* loss functions for user $u$ can be formulated as follows:

$$L_{nDCG}(u) = -\frac{\sum_{i=1}^{N}(2^{y_{ui}}-1)/\log_2(\tilde{R}_{ui}+1)}{\sum_{i=1}^{m_u} 1/\log_2(i+1)}, \quad (2.14)$$

$$L_{AP}(u) = -\frac{1}{m_u}\sum_{i=1}^{N}\frac{y_{ui}}{\tilde{R}_{ui}}\left(1 + \sum_{j=1\backslash i}^{N} y_{uj}\sigma(f_{uj}-f_{ui})\right), \quad (2.15)$$

$$L_{RR}(u) = -\sum_{i=1}^{N}\frac{y_{ui}}{\tilde{R}_{ui}}\prod_{j=1\backslash i}^{N}\left(1 - y_{uj}\sigma(f_{uj}-f_{ui})\right). \quad (2.16)$$

To the best of our knowledge, no method has been proposed yet for *RBP*-based listwise optimization. Inspired by the loss definitions for the other metrics, and again choosing to work with the normalized formulation of *nRBP*, we introduce the method for defining the corresponding loss function as follows.

By virtue of the monotonicity of the logarithm function, the recommendation model that optimizes $nRBP(u;p)$ also optimizes

$$\ln\left(\frac{nRBP(u;p)}{m_u}\right) = \ln\left(\frac{RBP(u;p)}{m_u}\right) - \ln\left(iRBP(u;p)\right). \quad (2.17)$$

Note that the second term is a constant for each user, so we focus on the first term. Based on Jensen's inequality, we can derive a lower bound for the first term as follows:

$$\begin{aligned}
\ln\left(\frac{RBP(u;p)}{m_u}\right) &= \ln(1-p) + \ln\left(\frac{1}{m_u}\sum_{i=1}^{N} y_{ui} p^{\tilde{R}_{ui}-1}\right) \\
&\geq \ln(1-p) + \frac{1}{m_u}\sum_{i=1}^{N}\ln\left(y_{ui} p^{\tilde{R}_{ui}-1}\right) \\
&= \ln(1-p) + \frac{1}{m_u}\sum_{i=1}^{N} y_{ui}(\tilde{R}_{ui}-1)\ln(p).
\end{aligned} \quad (2.18)$$

Note that the last equality holds because only $y_{ui}=1$ contributes to the summation, and the two remaining logarithmic terms are constant across users.

Because $p \in (0,1)$, $\ln(p)$ is negative, so maximizing the formulation in Eq. (2.18) becomes equivalent to minimizing $\sum_{i=1}^{N} y_{ui}(\tilde{R}_{ui}-1)$. In this way, our *RBP*-based optimization of the ranking treats all relevant items equally and aims at bringing them close to the top. Although convenient and intuitive, this function does not have common bounds across users. Therefore, if used alone as the optimization objective, it would make the training process sensitive to specific users. To (partially) resolve this issue, we come back to the second term in Eq. (2.17), find its own "lower bound" using Jensen's inequality and subtract it from Eq. (2.18) to make the lower bound equal 0 for all users. After dropping the logarithms, the regulated *nRBP* loss for user $u$ can now be denoted as

$$L_{nRBP}(u) = \sum_{i=1}^{N} y_{ui}(\tilde{R}_{ui}-1) - \sum_{j=1}^{m_u}(j-1). \quad (2.19)$$

In this way, the optimization of $nRBP$ becomes equivalent to an elegant direct optimization for the rank position of the relevant items. In view of the fact that the ideal situation leads to ranking all relevant items at the top, the listwise loss inspired by $RBP$ shows potential for achieving high ranking effectiveness across different evaluation criteria, making it an informative metric. Furthermore, $L_{nRBP}$ is independent of $p$, which makes it possible to conduct $RBP$-based optimization for different $p$ values in one unified framework. We note here once again that the regularization in Eq. (2.19) only has effect on the lower bound of the loss value range. The loss value remains user-sensitive and can still be arbitrarily large depending on the number of interactions $m_u$. We analyze the consequences of this in Section 2.5.3.

### 2.3.5. MODEL LEARNING

In recent years, Adaptive Moment Estimation (Adam) [57] has become one of the most popular optimizers. Compared to traditional optimizers like Stochastic Gradient Descent (SGD), its insensitivity to hyper-parameters and faster convergence makes it widely deployed in machine learning models. Despite these advantages, Adam tends to suffer from convergence and generalization power [149, 88]. Because in this chapter we investigate the generalization power of optimizing for different metrics, we still choose to optimize all our models with SGD.

## 2.4. EXPERIMENTAL DESIGN

Our goal is to investigate the capabilities of metrics used for optimization when the recommendation performance is assessed by the same or other metrics. In the following we explain the selection of datasets and experimental protocol for our experiments.

### 2.4.1. DATASETS

We selected four widely-used and real-world datasets to experiment with a diverse set of data. Two of them, CiteULike-a [128] and Epinions [120], contain unary data, while the other two, Sports & Outdoors and Home & Kitchen, are datasets with graded ratings from Amazon [93]. Amazon datasets contain integer relevance scores ranging from 1 to 5, so we need to binarize them before they can be used by our LTR methods. We choose to consider as positive only ratings of 4 and 5, which surely reflect a positive preference, and every other rating as a negative preference. In addition, and as is common in experimentation with recommenders, the absence of a rating is also taken as a negative interaction in all datasets [119]. As shown by Cañamares, Castells, and Moffat [23], although users with few ratings might exist in commercial services, they are usually filtered out in offline experiments because the lack of data leads to unreliable performance measurements. To address this issue, we only keep users with at least 25 relevant interactions in all datasets. The post-processed dataset statistics are shown in Table 2.1.

### 2.4.2. EXPERIMENTAL PROTOCOL

We use LensKit [37] to randomly split the data into training and test sets, stratifying by user: we sample 80% of their interactions for training, and leave the rest for testing. As a result, each user has at least 20 relevant interactions in the training set, and at least 5 in the test set. The evaluation metrics used are *RR*, *AP*, *nDCG* and *RBP* with *p* equal to 0.8, 0.9 and 0.95, so that we can assess recommendation performance under different degrees of user persistence. For pairwise optimization using LambdaRank, these six metrics are in line with six separate loss functions. In the listwise context, however, we have in total 4 loss functions because the loss for *nRBP* is independent of *p*. All models are trained with a considerable number of epochs (3,000), so that every metric has a more than reasonable chance to achieve its best performance. For each model, we select the epoch yielding the best performance on the corresponding evaluation metric. For the optimization of listwise *nRBP*, the optimal epoch is chosen for each of the 3 values of *p* separately, so that we actually have 3 different models.

Table 2.1: Dataset statistics.

| Dataset | #users | #items | #ratings | Density |
|---|---|---|---|---|
| CiteULike-a | 2,465 | 16,702 | 157,527 | 0.383% |
| Epinions | 4,690 | 32,592 | 325,154 | 0.213% |
| Sports & Outdoors | 9,123 | 119,404 | 342,311 | 0.031% |
| Home & Kitchen | 20,531 | 222,472 | 795,845 | 0.017% |

To reduce random error due to data splitting, we adopt a Monte Carlo cross-validation approach [36] and create three independent splits per dataset. Shi et al. [114] indicate that, for IR metrics that only rely on the ranking positions of relevant items, there is no need to consider all irrelevant items when training. As a result, a negative sampling process can significantly speed up training without hurting the overall performance. Negative sampling is, however, not only beneficial for efficiency. According to Cañamares and Castells [22], removing some (or even a significant number of) negative items from both the training and test sets can make the evaluation more informative and less biased with, balancing popularity and the average relevance of items across users. Such a strategy can therefore also make the evaluation more effective. We choose to inform the negative sampling process by the number of relevant items for each user, so we sample as many negative items as positives, twice as many, or five times as many; we denote this as the Negative Sampling Ratio (NSR). These negative items, along with the 80% of positives, form the complete training set for a user. In order to align the distributions of training and test sets, we complete the test set using the same approach: the remaining 20% of positives, plus 100%, 200% or 500% as many negatives.

In order to maximize the performance of the assessed models, we fine-tuned the learning rate of SGD and performed a search in the range {0.001, 0.01, 0.1} for LambdaRank and {0.001, 0.01, 0.1, 1, 3, 10} for listwise models. We also conducted a preliminary exploration on the number of latent factors for Matrix Factorization within the range {8, 16, 32, 64, 128}. The results showed that, although there is a positive correlation between the ranking effectiveness and the latent space dimensionality, such a correlation has no impact on the relative performance of different losses. Therefore, we do not analyze the effect of dimensionality in this chapter and simply fix the number of latent factors at 32 throughout the experiments.

We implement all models in PyTorch [96]. To accelerate the training process, we use CUDA and CuDNN on an NVIDIA GeForce GTX 1080Ti GPU.

## 2.5. RESULTS

In this section we compare the effectiveness of different metrics when used for optimization in ranking-based recommender. The goal of the analysis is threefold. First, in Section 2.5.1 we focus on the overall performance of pairwise and listwise models and investigate whether the practice of optimizing for the metric used in evaluation is justified in ranking-based recommendation. Then, in Section 2.5.2 we conduct a deeper analysis on the impact of a metric chosen for optimization on the ranking effectiveness assessed by different evaluation metrics. In doing so, we especially focus on the performance of the *RBP*-inspired objective functions introduced in this chapter. Finally, in Section 2.5.3 we investigate the effect of different metric optimization strategies on the recommendation utility for active and inactive users, with special emphasis on the impact of the missing upper bound of the *RBP*-based listwise loss function proposed in Eq. (2.19).

Due to space constraints, we do not report all the results obtained in our experiments. The reported results are, however, fully representative of the complete set of results that led to the final observations, conclusions and recommendations for future work.[1]

### 2.5.1. SHOULD WE OPTIMIZE THE METRIC USED TO EVALUATE?

Fig. 2.1 shows the performance of all pairwise and listwise learning objectives on all 6 evaluation metrics. While a negative correlation can be observed between recommendation effectiveness and the NSR, this does not necessarily mean that the models trained with more irrelevant items are worse. Because the training and test sets follow the same distribution, more negative items in the test set just make the relevance prediction task harder.

Several observations can be made from this figure. First, in both pairwise and listwise models, optimizing *RR* consistently yields significantly worse performance, even when the evaluation target is also *RR*. This observation supports previous findings that optimizing other metrics can achieve higher *RR* test scores than optimizing *RR* itself [35]. The explanation for this is two-fold. First, *RR* does not exploit all the information in the training data because it only focuses on the first relevant item, resulting in suboptimal models compared to optimizing other metrics. Second, as indicated by Webber et al. [135], *RR* is not well correlated with other informative metrics, such as *nDCG* and *AP*, which may optimize for *RR* but not the other way around. Because the performance gap of optimizing *RR* is stable and significant, we do not include it in further analysis and instead focus on the other 5 learning objectives (`metric_optim`), but still with all 6 evaluation metrics.

Second, listwise and pairwise methods behave differently when optimizing *nRBP*. In the pairwise context, the recommendation performance obtained by optimizing different *nRBP* losses is varied. Specifically, we find that optimizing with $p = 0.95$ outperforms optimizing with $p = 0.8$ or $p = 0.9$. This finding is consistent with prior research showing that *RBP*.95 is better correlated with informative metrics than with other different $p$'s [105, 86]. Such an inner-*RBP* advantage can also be explained by the nature of the metric. Because *p* models user persistence, a high value takes relevant items lying deeper in the list into account during training. This allows the model to account for more information, which benefits its performance. Although higher *p* also brings a slower weight decay, which does not favor prediction for the top of a recommendation list, this effect is insignificant for systems with binary relevance, where we do not need to put highly relevant items ahead of moderately relevant ones. In the listwise context, however, the results obtained when optimizing for *nRBP* with different *p* values are in general homogeneous. This shows that our *p*-independent *nRBP* loss provides a generic method to optimize for multiple *RBP* metrics. In this way, the concern of choosing a specific *p* for optimization is addressed in a simple and effective way.

---

[1]All data, code and full results are available at
https://github.com/roger-zhe-li/sigir21-newinsights.

Figure 2.1: Overall performance of pairwise and listwise methods. `metric_optim` denotes the metric-based losses for optimization, and a panel row denotes an evaluation metric. All results are averaged over the 3 data splits. Note that y-axes vary per row.

Last but not least, in both listwise and pairwise paradigms optimizing *nDCG*, *AP* and *RBP*-inspired losses achieves similar stable performance across different datasets, NSRs and evaluation targets. Such an observation, combined with the finding that optimizing *RR* leads to the worst ranking effectiveness when evaluating with *RR*, suggests that the practice to optimize and evaluate recommender systems with the same metric is not necessarily the best. A more detailed analysis on the relative advantages of individual metrics is given in the next section.

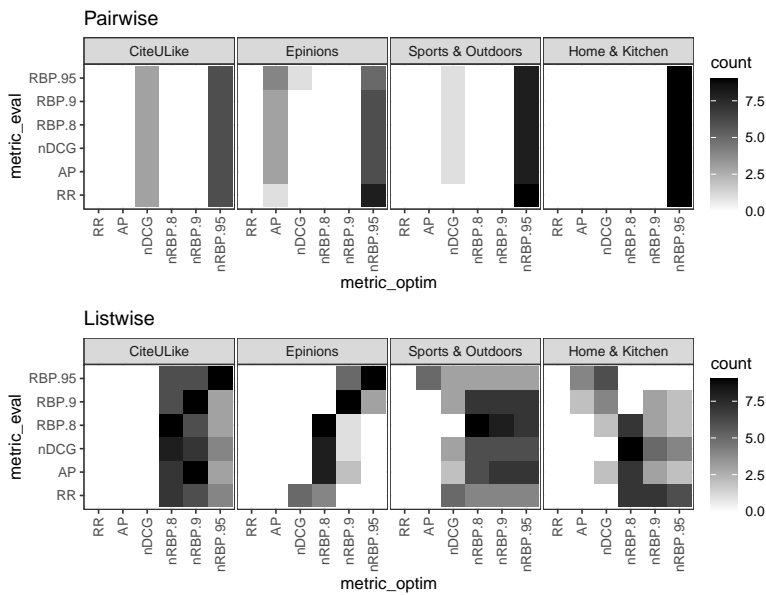

Figure 2.2: The frequency for each metric-based loss (`metric_optim`) of achieving the best performance on specific evaluation metrics (`metric_eval`), in all 3 data splits and all 3 NSR's.

### 2.5.2. IS *RBP* MORE EFFECTIVE AS OPTIMIZATION METRIC THAN OTHERS?

Fig. 2.2 shows how often each metric achieved the best test performance, across evaluation metrics, when used for optimization. Since frequencies are counted from 3 splits and 3 NSR's for each evaluation metric, the frequency is expected to have a row-wise sum of 9. Of course, it is possible for different losses to tie and obtain the best result for a certain case, so the row-wise sum is actually larger than 9 for several evaluation metrics, especially in the listwise case. Overall, we find that in both pairwise and listwise scenarios most of the best performance cases are achieved when optimizing *RBP*-inspired metrics. More importantly, in LambdaRank we even find that optimizing *nRBP*.95 shows a significant and clear advantage over all the other metrics.

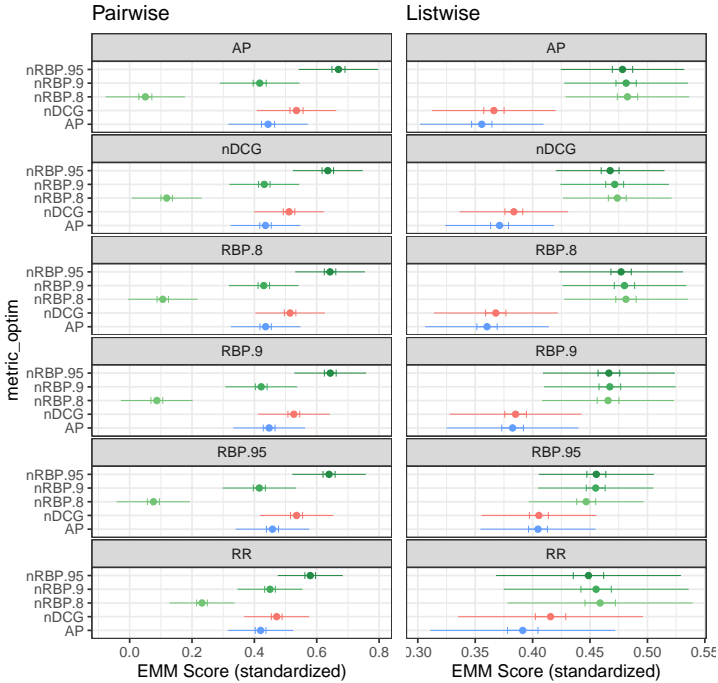

Figure 2.3: Estimated Marginal Mean (standardized) score for each `metric_optim`. Each panel represents an evaluation metric. Small bounded segments represent 95% confidence intervals. Long unbounded segments are prediction intervals.

Even though *RBP*-based losses seem to achieve the best performance, Fig. 2.1 suggests that differences could be too small or indistinguishable from random error, so next we proceed to a statistical analysis. First, we standardize performance scores within dataset-NSR combination to avoid homoscedasticity, because Fig. 2.1 evidences very different scales per dataset and NSR. This way, scores are comparable across metrics. We then fit a linear model on the standardized scores, using as independent variables the loss function, dataset and NSR, as well as their 2-factor interactions with loss function.[2] To properly compare the effect of each loss function while controlling for the other factors, we compute their Estimated Marginal Means (EMM) [108], as well as 95% confi-dence and prediction intervals, that is, what to expect *on average* over multiple training runs, and what to expect of an *individual* training run.

---

[2] The inclusion of dataset and NSR main effects does not inform the model in any way because of the standardization, but we keep them to follow the hierarchy principle of linear models.

Fig. 2.3 presents the EMM standardized performance scores of all metric-based losses except *RR*. These overall results show that, in LambdaRank, optimizing *nRBP*.95 achieves the best performance across all 6 evaluation metrics and shows a consistent and statistically significant advantage over the other losses. In listwise models, our *RBP*-inspired loss also achieves a statistically significant advantage over the others on all 6 evaluation metrics. These observations demonstrate the power of optimizing *nRBP*.

To further explore the stability of such performance gain across different datasets, we show in Fig. 2.4 the EMM scores, but faceted by dataset. Although the general superiority of *RBP*-based learning objectives is statistically significant on average, the overlapping prediction intervals indicate that it is not always necessarily the best option. In LambdaRank, the *nRBP*.95 loss performs best in all datasets except Epinions, where it is not statistically different from the *nDCG* and *AP* losses. Interestingly, we see that optimizing *nRBP*.8 lies in the opposite extreme and consistently achieves the worst performance. Even its prediction interval seldom overlaps with that of optimizing *nRBP*.9 or *nRBP*.95. In listwise recommendation, the *RBP*-inspired loss shows significant superiority over *AP* in a nearly consistent fashion, except for some notable cases like evaluating *RR* in the Epinions dataset or *RBP*.95 in the Home & Kitchen dataset. It also performs significantly better than *nDCG* in general, but both losses yield otherwise similar performance in several cases, especially in the Epinions and Sports & Outdoors datasets. On the CiteULike dataset, however, the advantage of *RBP*-based models is very clear and there is even no overlap between the prediction intervals, except when evaluating *RR*.

All in all, we draw the conclusion that optimizing *nRBP*.95 can help achieve recommendation effectiveness at least not worse than when optimizing other informative metrics, such as *nDCG* and *AP*. Furthermore, the performance of optimizing *nRBP* on different metrics is homogeneous, which means that our *RBP*-inspired listwise loss is able to help maximize *RBP* scores regardless of $p$. Moreover, according to the unreported results from our training logs, this homogeneity is not only achieved in evaluation scores, but also in the convergence process. When training with listwise *RBP*-inspired losses, we manage to get the optimal *RBP* values on all three $p$ values at a similar stage. In some models, the epochs to get all three optimal *RBP* scores are even the same, which means that the epoch with the optimal *RBP*.95 score also provides good scores on *RBP*.8 and *RBP*.9, and that we have the possibility to validate on only one $p$ value to satisfy different needs expressed by different values of $p$. Hence, our listwise *nRBP* optimization can serve as a generic choice for rank-based recommendation. We can explain this by the nature of our listwise *nRBP* loss. With our transformation in Eq. (2.19), we do not assign different weights to items ranked at different positions. Instead, we aim at bringing all relevant items to the top of the list, and treat all positive items as equally important. This provides the model with more abundant information to train.

The observations above point to the conclusion that, although the superiority brought by optimizing for *RBP*-based losses is not always significant and not fully consistent across datasets, we are still provided with a promising alternative metric to optimize in rank-based recommender systems. By optimizing *RBP*-based losses, we are able to get at least comparable performance as optimizing *nDCG* and *AP*, with very clear benefits in many cases. In the following section, we will conduct an analysis seeking the source of performance advantage of *RBP*-based listwise losses.

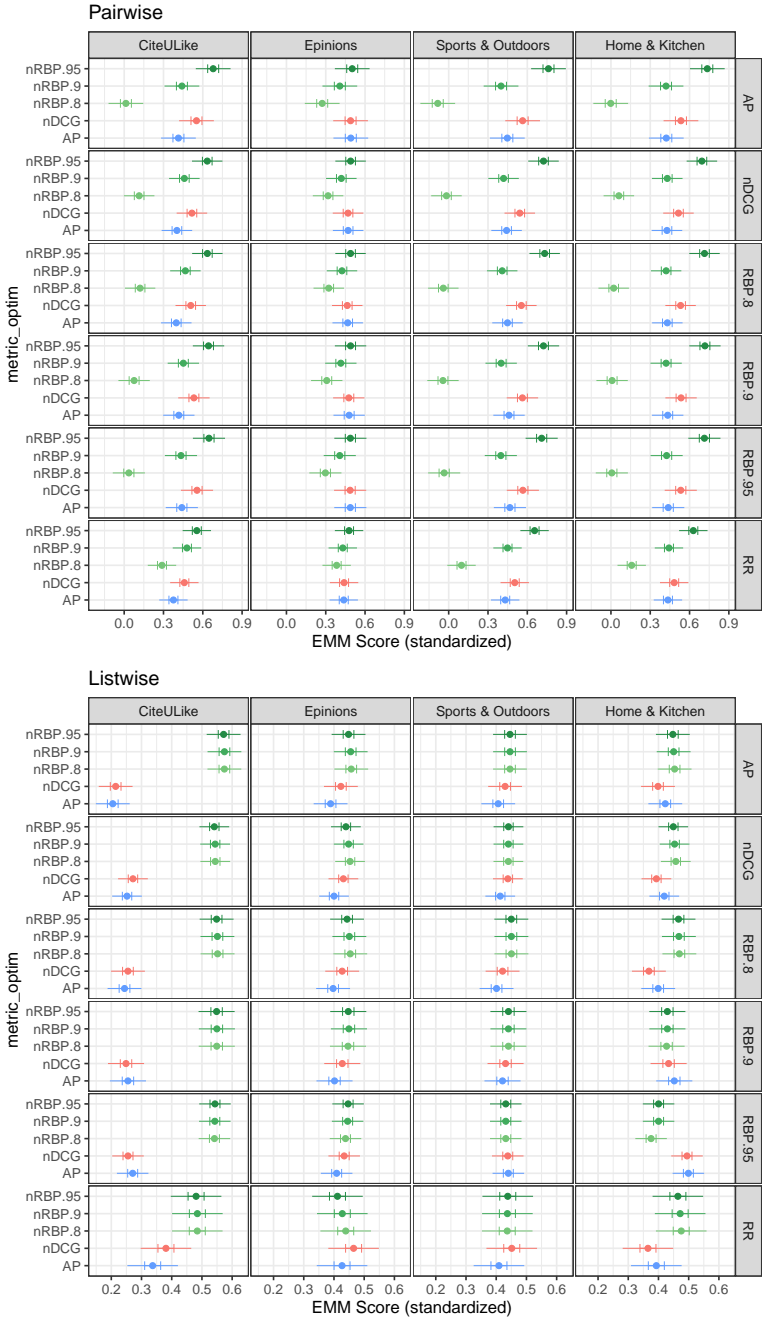

Figure 2.4: Estimated Marginal Mean (standardized) score for each `metric_optim`, by evaluation metric (rows) and dataset (columns). Small bounded segments represent 95% confidence intervals. Long unbounded segments are prediction intervals.

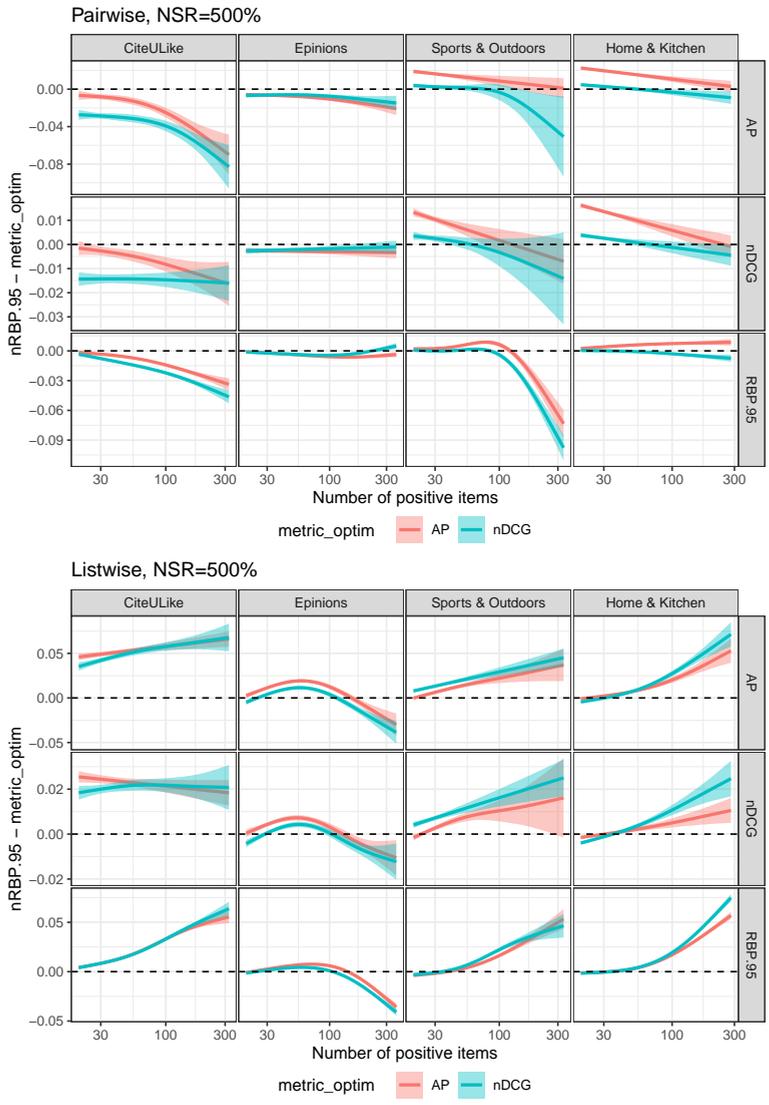

Figure 2.5: Score difference between optimizing *nRBP*.95 and *AP* or *nDCG* (higher is better for *nRBP*.95), when NSR=500%, as a function of the number of positive items for the user. Curves show spline-smoothed fits with 95% confidence intervals.

### 2.5.3. WHEN TO DEPLOY *RBP* FOR RECOMMENDATION?

The analysis in the previous section indicated that optimizing *nRBP* can provide recommendation performance comparable to that of *nDCG* and *AP* or even better. We may trace back the source of such effectiveness and identify the best way to deploy it if we analyze the properties of the *RBP*-based objective functions in the pairwise and listwise context. Similarly to the *nDCG* and *AP* losses, the pairwise *nRBP* loss in Eq. (2.10) guarantees strict bounds for the swap loss for each user, so that users are treated equally regardless of how many interactions they have. Contrary to this, the loss used for listwise *nRBP* in Eq. (2.19) is directly related to the predicted rank positions, which makes it have a different upper bound across users. Active users with more positive interactions are more likely to have larger loss values during training, especially for large *NSR*. Such an imbalance may benefit active users, but perhaps at the cost of sacrificing the utility for inactive users. We investigated whether this effect indeed occurs in our experiments.

Fig. 2.5 shows the performance difference between the *nRBP*.95 loss and *AP* or *nDCG* loss for every user, as a function of the number of positive items they contain in the training set. We deliberately choose to compare *nRBP*.95 because it achieves overall best results, and *AP* and *nDCG* for being the two next best losses, also with a properly bounded loss function. In addition, we only show results here for *NSR* = 500%, as it is expected to amplify the aforementioned bias, if any. We find that the *RBP*-inspired losses indeed perform differently under pairwise and listwise environments. In LambdaRank, where the *nRBP*.95 loss is also strictly bounded, we do not observe a clearly positive or negative correlation between the performance difference and the user activity level. In several cases, for active users with more items available for training, optimizing *nDCG* and *AP* is even more advantageous than optimizing *nRBP*.95. This means that active users did not benefit from the fully bounded *nRBP* loss, and there is no effectiveness imbalance between active and inactive users.

However, this observation does not hold in listwise models, as evidenced by the clear and significant positive correlation in all datasets except Epinions: active users do indeed benefit the most. More interestingly, this benefit for active users is not in detriment of the inactive users. As we can see in the figure, compared to optimizing *nDCG* and *AP*, optimizing the listwise *nRBP*.95 loss benefits all users on the CiteULike dataset, and achieves similar effectiveness for inactive users on the other 3 datasets. This superiority confirms our assumption that active users, whose listwise *nRBP* loss magnitudes are larger than for other users, can indeed get more effective recommendations due to a training process biased towards their utility, without negatively influencing the less active users. Such an insight is potentially interesting for some business applications of recommender systems, where it is beneficial to maximally serve loyal users without losing the stickiness of less active users.

## 2.6. CONCLUSION

Direct optimization of IR metrics has long been a hotspot in the research on ranking-based recommender systems. The intuitive and logical common practice is to build models by optimizing the same metric that will be used for evaluation. In this chapter, we reported the results of an extensive experimental study aiming at acquiring new insights about the strength of the foundations behind this practice and at learning more about what metric to optimize in order to maximize recommendation effectiveness. For this purpose, we expanded the scope of metrics usually deployed to define the objective functions for LTR approaches and focused on *RBP* as a promising alternative to other metrics such as *AP*, *nDCG* and *RR*.

Experimental evidence on both pairwise and listwise frameworks show that optimizing *AP*, *nDCG* and *nRBP* generally outperforms optimizing *RR*, and that optimizing *nRBP* is generally no less effective than optimizing *nDCG* or *AP*. These findings challenge the practice to optimize and evaluate ranking-based recommender systems using the same metric. Furthermore, the new generic listwise *RBP*-inspired loss proposed in this chapter was shown to be able to achieve the optimal performance for different values of the user persistence parameter, without the need to specify this parameter explicitly. Optimizing this loss even significantly outperformed the direct optimization of *nDCG* and *AP* in some cases, showing the high potential of *RBP* for developing ranking-based recommender systems. Finally, and due to the lack of a common upper bound across users, our proposed listwise *nRBP* loss benefits active users more than *nDCG* and *AP*, but without hurting the effectiveness for inactive users. This makes optimization of the proposed *RBP*-based listwise loss interesting for some business application cases favoring loyal users.

For future work, we will experiment with more advanced recommendation models and larger datasets to study the extent to which our conclusions and insights generalize to other settings. We will also analyze to what extent the exclusion of very inactive users affects our conclusions, especially with regards to the bounds of *nRBP* losses. Furthermore, it would be interesting to theoretically investigate the source of the effectiveness of *RBP* even deeper. Our results show that it is a promising metric to optimize when learning to rank for recommendation, pointing to the possibility of finding even more IR metrics that could show similar potential.

# 3

# LEAVE NO USER BEHIND: TOWARDS IMPROVING THE UTILITY OF RECOMMENDER SYSTEMS FOR NON-MAINSTREAM USERS

## 3.1. INTRODUCTION

Collaborative Filtering (CF) models are the most investigated and deployed models in the domain of recommender systems [4]. These models assume that users who had similar preferences on items of a specific kind (e.g. books, movies) in the past may continue having similar preferences on other items of the same kind. The preferences of the users are expressed through their explicit (e.g. ratings) or implicit (e.g. clicks, downloads) interactions with items.

Among the CF models, *Matrix Factorization* (MF) [60], which tries to find a representation of both users and items in the same latent factor space, has long been the most successful and most widely deployed CF model. More recently, generalized factorization models, such as factorization machines [99], have been proposed, exploiting input beyond user-item interactions to learn the latent space. Exploiting more input, such as contextual features and other types of useful side information about users and items, was shown to further improve the recommendation quality. The potential for further improvement, for instance by relying on more abundant input data including audio, visual and textual item descriptions or social network dynamics, came only when deep neural networks (DNN) [74] entered the recommendation domain and enabled more sophisticated user/item representation space learning. In particular, textual data acquired from websites have been extensively exploited for this purpose, allowing users to leave review comments for items along with ratings. For this type of data, DNN-based user/item modeling utilizing NLP techniques has been shown to achieve significantly higher recommendation performance [146, 31, 122, 75] as well as provide convincing explanations [26, 32, 131, 95].

While these developments have greatly contributed to the improvement of the overall recommendation accuracy, one problem has remained largely unsolved, namely the presence of various types of biases in the learned recommendation models. In this chapter we focus on the bias towards the so-called *mainstream users*. A mainstream user often prefers items liked by many people and also reacts negatively to items widely disliked by others [106]. Contrary to this, non-mainstream users typically show interest on rarely-visited items or have an opposite attitude towards widely accepted or rejected items. Such a "grey sheep" property [148] makes these users different from others, making it difficult for a CF algorithm to identify similar peers. This leads to recommendations of a generally lower quality, because recommendations for these users are based on neighbors with insufficiently similar preferences. Furthermore, non-mainstream users are typically a minority and the numerous consistent user-item interactions within the cluster of mainstream users are likely to be dominant in steering the process of learning the user/item representation space. Because of this, the non-mainstream users and their preferred ("outlier") items become underrepresented in such a space, leading to inequality of the recommendation performance across the user population. This is the *mainstream bias*, the tendency to provide better recommendations to the mainstream users. Such bias could make non-mainstream users draw insufficient utility from a recommender system and could discourage them from using it anymore. This could lead to online businesses starting to lose customers. For the information and news recommender systems, however, we foresee even more serious consequences. Recommender systems may namely become less inclusive with respect to non-mainstream opinions and (e.g. political) views and in this way contribute to undesired long-term effects, like intellectual segregation and societal polarization.

In this chapter we build on the success of DNNs for developing recommendation models and propose a simple but effective solution towards neutralizing the mainstream bias. With our new recommendation model, referred to as **N**eural **A**uto**E**ncoder **C**ollaborative **F**iltering (NAECF), we introduce adversarial conditions to the process of learning the recommendation algorithm, which in this specific case is realized as the minimization of the rating prediction error. The adversarial conditions are imposed by *autoencoders* [13], a deep learning architecture widely used for recommendation [109, 94, 77], added to a state-of-the-art DNN-based recommendation framework. They enforce that the user and item representations are learned in a way such that they preserve their specific and unique properties before being fed to the rating predictor.

Since this preservation is achieved for all users, mainstream or not, the autoencoders prevent that the learned representations are biased towards the users with a mainstream taste. The results of experiments conducted on different domains and scales of the real-world datasets from Amazon [80] show that the representations learned in this way indeed help to de-bias the produced recommendations (predicted ratings in this case). Compared to the case without deploying the adversarial conditions, our proposed method produces significantly better recommendations for non-mainstream users while largely maintaining the recommendation quality for mainstream users. We clearly show that this performance improvement largely stems from adding adversarial conditions to the process of user and item representation learning. In addition, our experiments demonstrate the benefit the non-mainstream users draw from the application of content-based features such as online reviews, further highlighting their value for achieving high recommendation quality across the user community.

The proposed NAECF approach is, to the best of our knowledge, the first to enforce preserving the unique user and item properties as the adversary to the process of learning how to recommend. This allows us to effectively address the mainstream bias in recommender systems, which has not been extensively studied this far.

A recommender system can be designed to either predict ratings or rank items. Despite the latter is picking up momentum in the field, we still choose in this chapter to follow the rating prediction paradigm. The main reason for this lies in the core of our contribution, which is to investigate how a state-of-the-art recommendation framework may be extended in order to de-bias the process of generating recommendations. Since the framework we build upon was evaluated in terms of rating prediction, we follow this same paradigm in this chapter. Nonetheless, the user-item representation space generated by the autoencoders can serve to predict both ranking and ratings, so we do not consider our choice to limit the broad application of our proposal to the recommendation practice.

The results of this chapter can be fully reproduced with data and code available online[1].

---

[1]https://github.com/roger-zhe-li/wsdm21-mainstream

## 3.2. RELATED WORK

Our work relates mainly to two topics: biases in recommender systems and review-based user/item modeling.

### 3.2.1. BIASES IN RECOMMENDER SYSTEMS

Potential biases in the training data have already been recognized in early work on matrix factorization for recommendation. Koren, Bell, and Volinsky [60] introduced a correction in the dot-product rating prediction formula to incorporate rating biases across users, that is, how the rating scale is interpreted by different users. Another bias related to ratings is the anchoring bias; it emerges from the influence of previous recommendations to a user on that user's future ratings. Adomavicius et al. [5] explored two approaches to neutralize this bias. The first approach involves computational post-hoc adjustments of the ratings that are known to be biased. The second approach involves a user interface by which the system tries to prevent this bias *during* rating. A different sort of bias is the popularity bias, due to which popular items may be recommended more frequently than other, less popular items (e.g., long-tail). Abdollahpouri, Burke, and Mobasher [1] proposed an add-on to a general collaborative filtering algorithm by which a trade-off between accuracy and long-tail coverage can be tuned. More recently, the discussion about biases has increasingly been conducted in the context of resolving ethical and societal issues when deploying recommender systems in practice, such as polarization [98], fairness [82] and discrimination [78], giving a further boost to the research on this topic.

In this chapter we focus on the aforementioned mainstream bias. While being conceptually close to popularity bias [2, 62], there is an important difference between the two. Popularity bias could lead to a separation between more and less popular items, similar to the separation of items into those being interesting to mainstream and non-mainstream users. However, popularity bias is not informative regarding the way a recommender system serves different groups of users. According to Steck [118], users may tend to provide feedback on popular items simply by following (being influenced by) other users. In this way, their preferences are likely to be unconsciously driven away from their real interest. By focusing on mainstreamness, we explicitly look at the bias in the user population.

Kowald, Schedl, and Lex [62] demonstrated that non-mainstream music listeners are likely to receive the worst recommendations. Schedl and Bauer [107] investigated music preferences across age groups. They observed that, although only taking a small proportion of users, kids and adolescents have significantly different preferences from other age groups in terms of music genres, and the recommendation performance on these two groups is also distinctive among all users. To repair the unfairness caused by the mainstream bias, several recent works aim at identifying non-mainstream music listeners and using the power of cultural aspects [106, 89] and human memories [61] to better profile these underrepresented users in recommender systems. Despite the reported progress, existing methods to alleviate the mainstream bias usually rely on their specific definitions of mainstreamness, which may limit the findings. Furthermore, these methods tend to split users into different mainstreamness groups for individual training. This setting may cause the loss of recommendation accuracy due to not exploiting the cross-group collaborative information. The approach proposed in this chapter aims at neutralizing the mainstream bias in a more generic fashion and without divisions within the user population.

### 3.2.2. REVIEW-BASED USER/ITEM MODELING

Supported by the rapid development of natural language processing (NLP) techniques, online reviews have increasingly been identified as an important source of useful information for addressing data sparsity issues in recommendation. Exploiting these reviews has led to several advanced recommendation concepts, pioneered by Collaborative Deep Learning (CDL) [129]. This concept introduces a hierarchical Bayesian model using Stacked Denoising Autoencoders (SDAE) to reconstruct the rating matrix from encoded textual reviews. Another method, DeepCoNN [146], unifies the processes of learning the user/item representation and rating prediction in an end-to-end model. The unification is achieved through a combination of Convolutional Neural Networks (CNN) and factorization machines. Due to the sequential nature of reviews, Recurrent Neural Networks (RNN) and attention models are also widely used for user and item feature learning. Wu et al. [138] trained the review representations and ratings jointly within a Long Short-Term Memory (LSTM) framework for movie recommendation. Chen et al. [26] extended the DeepCoNN concept by incorporating attention factors into NARRE, a DeepCoNN-based framework, to provide convincing explanations. MPCN [122] is another attention-based model, which uses two hierarchical attention layers to infer the review importance. Although the use of text reviews has partially resolved the data sparsity issues, a more direct way to achieve this is to increase the scale of the training data. As an example, AugCF [132] was proposed on top of DeepCoNN to augment review and rating data using Generative Adversarial Networks [42]. All models mentioned above represent users and items following the same principle, and the representations are derived from the same data source. Contrary to this, NPA [140] and NeuHash-CF [44] represent users and items in different ways. While they model the items using content-based information, the users are represented by one-hot coded user ID.

Despite the remarks expressed in literature state that reviews serve the recommendation better as regularizers than features [104], the models mentioned above have been reported to achieve remarkable overall recommendation accuracy, showing the benefit of using textual review data as input. In this chapter, we look at online reviews from a different angle and further than accuracy alone. We analyze their value in achieving better user representations that allow us to balance the recommendation quality across users. We show that, with our proposed recommendation model, reviews can be instrumental in neutralizing the mainstream bias.

## 3.3. PROPOSED MODEL: NAECF

The architecture of the proposed NAECF model is illustrated in Fig. 3.1. The scheme shows that with NAECF we pursue two learning goals simultaneously: maximizing the recommendation accuracy and reconstructing the users' and items' original feature vectors in the autoencoders. These feature vectors consist of the texts of user reviews, so we refer to the process taking place in the two AEs as "text reconstruction". Recommenda-tion accuracy may be achieved by optimizing for rating prediction or ranking prediction. Since DeepCoNN [146], the strongest baseline for comparison, is designed for rating prediction, we also take rating prediction as the criterion for recommendation optimization. This allows us to assess specifically the effect of enforcing user and item reconstruction as an adversarial condition to recommendation optimization on the mainstream bias. If the effect is there, it can also be expected if a ranking prediction scheme is expanded in the same way.

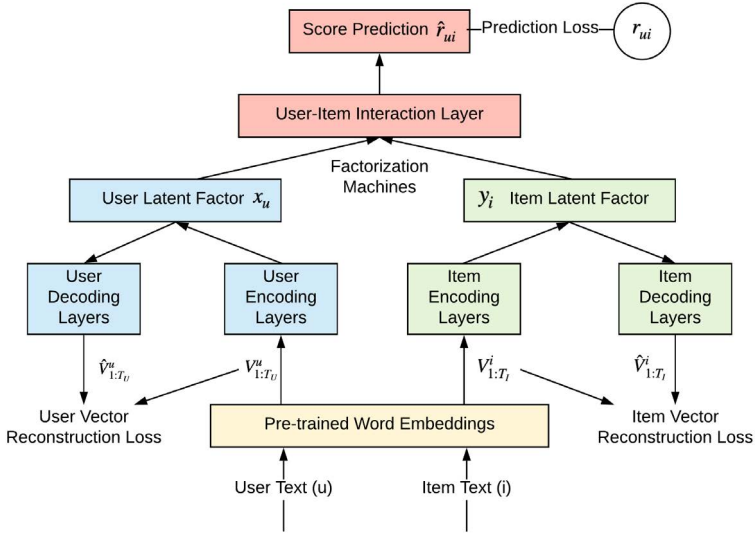

Figure 3.1: Overall architecture of NAECF.

### 3.3.1. MODEL FORMULATION

The data we use consist of tuples $(u, i, r_{ui}, c_{ui})$, representing a user $u$ providing a rating $r_{ui}$ to item $i$ and leaving a review text $c_{ui}$ for said item. Based on Fig. 3.1, we see the realization of the overall goal of NAECF by minimizing the following loss:

$$L = L_R + w(L_U + L_I),  \qquad (3.1)$$

where $L_R$, $L_U$ and $L_I$ are, respectively, the mean rating prediction loss, and the mean text reconstruction losses for users and items. The constant $w$ is a weight determining the relative influence of user and item AEs compared to the rating prediction module. The three losses are defined by the following expressions:

$$L_R = \frac{1}{N_R} \sum_{u,i} loss_R(u, i) \qquad (3.2)$$

$$L_U = \frac{1}{N_U} \sum_u loss_U(u) \qquad (3.3)$$

$$L_I = \frac{1}{N_I} \sum_i loss_I(i), \qquad (3.4)$$

where $N_R$, $N_U$ and $N_I$ represent the number of interactions, users and items in the training set, respectively. Normalizing by these terms makes the effect of the weight $w$ invariant to the statistics of the dataset.

### 3.3.2. LEARNING FOR RATING PREDICTION[2]

In NAECF, the rating prediction loss for an individual user-item interaction is computed as a traditional squared loss

$$loss_R(u,i) = \left(\frac{r_{ui} - \hat{r}_{ui}}{r_{max} - r_{min}}\right)^2, \tag{3.5}$$

where $\hat{r}_{ui}$ is the predicted rating given by user $u$ to item $i$. The loss is normalized by the limits of the rating scale used in the dataset, so that $L_R$ is bounded between 0 and 1. The prediction is computed for user and item representations $\boldsymbol{x}_u$ and $\boldsymbol{y}_i$, which encode the text features as low-rank latent vectors for users and items, respectively. Specifically, they are extracted as the bottlenecks of the corresponding AEs, as indicated by the green and blue blocks in Fig. 3.1. We follow the settings of DeepCoNN [146] with a Factorization Machine layer [100], and compute rating prediction as

$$\hat{r}_{ui} = \hat{a}_0 + \sum_{m=1}^{|\hat{z}|} \hat{a}_m \hat{z}_m + \sum_{m=1}^{|\hat{z}|} \sum_{n=m+1}^{|\hat{z}|} \langle \hat{\boldsymbol{v}}_m, \hat{\boldsymbol{v}}_n \rangle \hat{z}_m \hat{z}_n, \tag{3.6}$$

where $\hat{a}_0$ denotes the global bias, $\hat{z}$ denotes the concatenation of user and item vectors used in FM, and $\hat{a}_m$ denotes the importance of latent factor $\hat{z}_m$. Second order interactions are modeled by $\langle \hat{\boldsymbol{v}}_m, \hat{\boldsymbol{v}}_n \rangle = \sum_{f=1}^{|\hat{z}|} \hat{v}_{m,f} \hat{v}_{n,f}$.

### 3.3.3. LEARNING FOR TEXTUAL FEATURE TRANSFORMATION AND RECONSTRUCTION

The latent factors $\boldsymbol{x}_u$ and $\boldsymbol{y}_i$ are used not only for rating prediction, as indicated in the previous section, but also to reconstruct the original user and item representations in the computation of text reconstruction losses. We use an encoder to generate latent factors, which takes an initial user representation $V_u$ or item representation $V_i$. We deploy the strategy proposed by DeepCoNN [146], that applies TextCNNs [56] for feature transformation. For an arbitrary user $u$, we extract all review texts they authored and concatenate them into a single long document. Similar to the top NLP models like BERT [34] and GPT-2 [97], here we adopt a cutoff length $T_U$ to truncate words exceeding the limit. For users with fewer than $T_U$ words, we pad empty words, denoted by <UNK> and initialized by zeros. This way all users are represented by the same number of review tokens. Then we introduce a look-up layer to get the initial individual word embeddings from a pretrained model. By concatenating them, we obtain user embedding $V_{1:T_U}^u$. Similarly, we obtain item embedding $V_{1:T_I}^i$ for item $i$.

Encoding these initial $V_{1:T_U}^u$ and $V_{1:T_I}^i$ embeddings results in the latent factors, which then serve as input to a decoder that we introduce to create reconstructed embeddings $\hat{V}^u_{1:T_U}$ and $\hat{V}^i_{1:T_I}$. The architecture of the decoder is symmetric to the encoder with deconvolution and unpooling layers, as shown in Fig. 3.2. All hyper-parameters used in the decoding stage are the same as in the encoding stage.

---

[2] Compared to the original published paper, this subsection is slightly modified for better clarity.

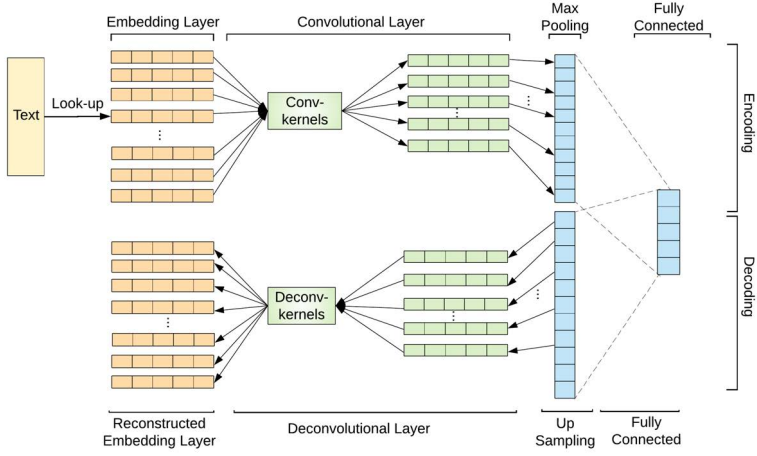

Figure 3.2: Architecture of the convolutional autoencoder for text feature transformation and extraction.

The success of reconstructing initial user and item embeddings is modeled by the text reconstruction losses, which are computed for each user and item. In order to have scores on a bounded scale, we rely on the cosine similarity to measure text reconstruction loss. Unlike most cases in text analysis where embedding values are positive, the original pre-trained embeddings we use in this chapter do have negative values, making the cosine similarity range from -1 to 1. Therefore, we also normalize cosine similarities so that the scales of $L_U$ and $L_I$ are comparable to that of $L_R$. This leads to the following formulation of the individual text reconstruction losses:

$$loss_U(u) = \left(\frac{1 - cos\left(V^u_{1:T_U}, \hat{V}^u_{1:T_U}\right)}{2}\right)^2, \tag{3.7}$$

$$loss_I(i) = \left(\frac{1 - cos\left(V^i_{1:T_I}, \hat{V}^i_{1:T_I}\right)}{2}\right)^2, \tag{3.8}$$

where $cos$ stands for the cosine similarity between the original vectors and the reconstructed ones.

### 3.3.4. MODEL LEARNING

We use Adaptive Moment Estimation (Adam) [57] to minimize the overall loss function in Eq. (3.1). This way, training converges fast and the learning rate is adapted during the process.

Table 3.1: Statistics of the datasets.

| Dataset | #users | #items | #ratings | Sparsity | #words |
|---|---|---|---|---|---|
| Instant videos | 5,130 | 1,685 | 37,126 | 99.57 | 19M |
| Digital music | 5,541 | 3,568 | 64,706 | 99.67 | 73M |
| BeerAdvocate | 3,703 | 37,580 | 393,035 | 99.72 | 198M |

## 3.4. EXPERIMENTAL DESIGN

Here we present a series of experiments designed to evaluate the proposed NAECF model through the following research questions:

- RQ1: Does NAECF improve the recommendation for non-mainstream users, creating a better balance across users?

- RQ2: What is the effect of using reviews and textual feature transformations on mainstream and non-mainstream users?

- RQ3: What is the correlation between recommendation accuracy and the difficulty of user feature reconstruction?

### 3.4.1. DATA AND METRICS

In this chapter we focus on improving the recommendation for non-mainstream users, and investigate the power of text reviews for this purpose. Therefore, the selected datasets are all review-based (see Table 3.1). We use two Amazon real-world datasets[3] covering different recommendation domains, namely instant videos and digital music, and another dataset from BeerAdvocate [79][4]. The ratings all range from 1 to 5. However, in the Amazon datasets ratings are integers, while in the BeerAdvocate dataset they are multiples of 0.5. Users in the Amazon datasets have at least 5 interactions. To align with this setting, we filter the BeerAdvocate dataset using the same threshold. Due to the unavailability of computational resources, we randomly sampled 25% of users to form a BeerAdvocate subset.

Following the original setting of DeepCoNN [146] and its latest related research [104], we use the Google News pre-trained word vectors [84] to generate pre-trained word embeddings. Each word in the review is thus represented as a 300-dimension vector.

We evaluate the rating prediction accuracy by computing the conventional Root-Mean-Square Error on the test set:

$$rRMSE = \sqrt{\frac{\sum_{u,i}(r_{ui} - \hat{r}_{ui})^2}{N}}. \tag{3.9}$$

where $N$ is the number of ratings. To evaluate recommendation performance for individual users, we also report per-user RMSE ($uRMSE$, as opposed to $rRMSE$) for further investigation. We cap the predicted ratings to [1,5], so there are no out-of-bounds values.

---
[3] http://jmcauley.ucsd.edu/data/amazon/
[4] http://snap.stanford.edu/data/web-BeerAdvocate.html

### 3.4.2. BASELINES

We compare the performance of our proposed NAECF model with two related recommendation models:

- **Matrix Factorization [60].** We use MF as a classical, pure similarity-based CF baseline. All non-textual hyper-parameters in NAECF are reused.

- **DeepCoNN [146].** This is the pioneering work and state-of-the-art method that introduces deep learning techniques to build a text-based recommender system. User and item features are extracted in parallel, and their interaction is realized by means of factorization machines (FM). Although there are other text-based models following a similar architecture, such as NARRE [26] and [132], that may outperform DeepCoNN, the components they added for better recommendation performance are mainly attention layers or data augmentation modules, introducing no significant change in the model architecture. Therefore, to focus on the effect of autoencoders in NAECF, we still adopt DeepCoNN as the strongest and most relevant baseline.

### 3.4.3. EXPERIMENTAL PROTOCOL

We randomly split the datasets into training, validation and test sets with proportions 80%, 10% and 10%, respectively. To address the influence of the data splitting strategy, we set 10 different random seeds and thus use 10 different splits. While all users have at least 5 interactions in total, a random split may distribute these interactions unevenly across sets, such that there may be users with only one rating in the training set. To address this potentially unreliable situation, we only account for users with at least 3 interactions in the training set for evaluation.

We first do a grid search on the two Amazon datasets separately to fix the hyperparameters on DeepCoNN. Then we reuse them for the investigation of NAECF. The hyper-parameters tuned are listed below, with the optimal values indicated in bold:

- Number of latent factors for DeepCoNN and NAECF: $\{5, 10, 20, 50, \mathbf{100}, 200, 500\}$. All latent factors are initialized with a Uniform distribution between $-0.01$ and $0.01$.

- Learning rate: $\{\mathbf{0.00001}, 0.0001, 0.001, 0.01, 0.03, 0.1\}$.

- Dropout rate to avoid overfitting: $\{0, 0.1, \mathbf{0.2}, 0.5\}$.

- Batch size: $\{32, 64, 128, \mathbf{256}, 512, 1024\}$.

- Number of words: $\{128, 256, \mathbf{512}, 1024, 2048\}$.

- Length of CNN kernels: $\{2, \mathbf{3}, 4\}$.

Using the DeepCoNN architecture as reference, we investigate the impact of text reconstruction loss with different weights. Since our main concern in this chapter is the effect of adding adversarial conditions via AEs to the original DeepCoNN setting, we weight the user and item AEs with the same weight $w$, as shown in Eq. (3.1). Specifically, we consider weight values in the set $\{0, 0.1, 0.2, 0.5, 1, 2, 5, 10\}$. Note that NAECF reduces to DeepCoNN when $w = 0$. Similar to the fine-tuning of the hyper-parameters, the optimal weight is selected on the validation set.

Autoencoders act as adversaries to the rating prediction process, so their activation may lower the overall validation $rRMSE$. Therefore, we deploy a two-stage training strategy: we set $w = 0$ in the first 50 epochs as a pre-training process to get the model ready to train for NAECF, and then change $w$ to the value we are tuning for the next 50 epochs.

Since NAECF does not chase the best overall performance, but rather a better balance across users, we follow a different validation strategy for $w$. First, we separate users in bins based on their $uRMSE$ score with DeepCoNN; to stress more on the performance for non-mainstream users, we use the 4 uneven bins defined by percentiles 10, 50 and 90 of the $uRMSE$ distribution. The performance gain with respect to DeepCoNN is then computed using these bins as strata, assigning smaller importance to the first and last bins, that is, users with a good recommendation and users who are extremely difficult to model. This way, the assessment of model capability is better aligned with our purposes. Gain is thus defined as follows:

$$\Delta = 0.1\Delta_1 + 0.4\Delta_2 + 0.4\Delta_3 + 0.1\Delta_4, \qquad (3.10)$$

where $\Delta_b$ indicates the mean $uRMSE$ difference between DeepCoNN and NAECF in user bin $b$, and bin weights reflect the fraction of users they contain out of the total sample. A positive $\Delta$ value means NAECF improves upon DeepCoNN.

All models are implemented in PyTorch [96], with CUDA and CuDNN for acceleration on an NVIDIA GeForce GTX 1080Ti GPU.

## 3.5. RESULTS

In this section, we present and analyze the experimental results. As a summary, Table 3.2 presents the mean performance of all models over the 10 splits per dataset. It can be seen that DeepCoNN and NAECFs show significantly better recommendation accuracy than MF (paired $t$-test, $p < 0.05$ [125]), and that NAECF and DeepCoNN perform similarly overall, provided that the weight of the text reconstruction loss is not too high. Furthermore, in Section 3.5.1 we show that NAECF, while maintaining similar overall recommendation quality as DeepCoNN, manages to create a significantly better balance across users thanks to the introduction of the user and item reconstruction losses as adversaries to the rating prediction optimization. Finally, in Section 3.5.2 we dive deeper into the ability of the autocorrelates to reconstruct users from the learned representations, and how this correlates with the recommendation performance per user. This analysis sheds more light on the mechanics underlying NAECF and the reported results.

Table 3.2: $rRMSE$ over 10 data splits for all recommendation models in all three datasets ($mean \pm std.dev.$). Bold for best results per dataset. Red for statistically significant loss with respect to the best ($t$-test, $p < 0.05$).

| Dataset | MF | DeepCoNN | NAECF | | | | | | |
|---|---|---|---|---|---|---|---|---|---|
| | | | $w = 0.1$ | $w = 0.2$ | $w = 0.5$ | $w = 1.0$ | $w = 2.0$ | $w = 5.0$ | $w = 10.0$ |
| Instant Video | 1.1600 ±.0264 | 0.9744 ±.0145 | **0.9732** ±**.0149** | 0.9749 ±.0122 | 0.9754 ±.0159 | 0.9757 ±.0169 | 0.9798 ±.0212 | 0.9896 ±.0221 | 0.9967 ±.0221 |
| Digital Music | 1.0466 ±.0097 | **0.9078** ±**.0138** | 0.9083 ±.0115 | 0.9106 ±.0128 | 0.9097 ±.0114 | 0.9104 ±.0128 | 0.9118 ±.0134 | 0.9167 ±.0108 | 0.9219 ±.0146 |
| BeerAdvocate | 1.0442 ±.0048 | 0.6722 ±.0090 | 0.6707 ±.0064 | **0.6692** ±**.0035** | 0.6746 ±.0059 | 0.6756 ±.0082 | 0.6785 ±.0098 | 0.6899 ±.0137 | 0.7068 ±.0278 |

Table 3.3: Weights $w$ yielding the best performance gain per split on the validation set.

| Dataset | Split | | | | | | | | | |
|---|---|---|---|---|---|---|---|---|---|---|
| | 1 | 2 | 3 | 4 | 5 | 6 | 7 | 8 | 9 | 10 |
| Instant Video | 2 | 5 | 0.1 | 10 | 1 | 5 | 0.1 | 0.1 | 0 | 0.5 |
| Digital Music | 0 | 0 | 5 | 5 | 2 | 0.1 | 2 | 0.1 | 0.1 | 0.5 |
| BeerAdvocate | 0.1 | 0.2 | 0 | 0.5 | 0 | 0.2 | 0.5 | 0.1 | 0.1 | 0.2 |

### 3.5.1. PERFORMANCE BALANCE ACROSS USERS

In order to answer research questions RQ1 and RQ2, we investigate the effect of autoencoders and text reviews on the recommendations for non-mainstream and mainstream users.

**EFFECT OF AUTOENCODERS.**

As an adversarial learning model, NAECF has two conflicting goals: minimizing the text reconstruction losses $L_U$ and $L_I$ versus minimizing the rating prediction loss $L_R$. If the weight of the text reconstruction loss is too small, autoencoders cannot exert sufficient influence on the training process, making them ineffective regarding the mainstream bias. Conversely, if the text reconstruction loss dominates the training process, we expect to have a significant drop in terms of overall rating prediction accuracy. Following the validation process in Section 3.4.3, we chose the weight $w$ with the best gain $\Delta$ on the validation set as the optimal one.

Table 3.3 reports the optimal validation-set weight per split. As the table shows, in 5 of the 30 splits a weight $w = 0$ achieved the best gain; note that such cases correspond to a simple re-training of DeepCoNN. However, in the vast majority of cases a weight different from zero yielded a better performance gain, although there does not appear to be a single optimal weight for the autoencoders in NAECF. Overall, this suggests that $w$ is a hyperparameter to tune on a case-by-case basis, and that autoencoders are expected to help when the characteristics of the data allow for it; sometimes they do not lead to a substantial gain over DeepCoNN. Furthermore, and based on detailed $rRMSE$ results not reported in the chapter, we see that the weights with the best performance gains often lead to lower overall performance (7, 6, and 6 out of 10 seeds in three datasets). This contrast shows that a high score on an overall accuracy metric like $rRMSE$ does not necessarily reflect a good balance across individual users.

After the optimal weights are chosen on the validation set, we turn our attention to the corresponding test-set results. In Table 3.4 we report the average performance gains of NAECF over DeepCoNN, both per bin and overall. The table shows that users in the central bins (ie. central 80% of users) receive a statistically significant performance gain on all datasets, which is exactly the users that we specifically target in NAECF. For the two Amazon datasets, these are also the bins receiving the largest gains; for the BeerAdvocate dataset it is the first bin that has the highest gain, though the difference is not significant from the second bin. In fact, the gains and losses observed for the 10% of users in the first bin are not statistically different from zero, which means that users that already receive good performance are neither helped nor punished by NAECF. Therefore, the application of autoencoders as adversaries to the rating prediction problem does not sacrifice performance for the mainstream users. Finally, we observe that the 10% of users in the last bin do receive a statistically significant performance loss. While unfortunate, such loss is a collateral damage on a minority of users who are hard to satisfy anyway, in benefit of the bulk of users who now receive better recommendations. Averaging gains across bins, as indicated in Eq. (3.10), we see that NAECF yields statistically better results than DeepCoNN on all datasets. This indicates the overall success of NAECF to create a better balance across users. In general, we help most of the non-mainstream users without hurting mainstream users.

Table 3.4: Test-set performance gains averaged over splits (higher is better): per-bin gain $\Delta_b$ and overall gain $\Delta$. Green/red for gains/losses statistically different from 0 ($t$-test, $p < 0.05$).

| Dataset | $\Delta_1$ | $\Delta_2$ | $\Delta_3$ | $\Delta_4$ | $\Delta$ |
|---|---|---|---|---|---|
| Instant Video | -0.0035 | 0.0256 | 0.0267 | -0.0308 | 0.0175 |
| Digital Music | 0.0036 | 0.0184 | 0.0106 | -0.0167 | 0.0103 |
| BeerAdvocate | 0.0119 | 0.0117 | 0.0063 | -0.0115 | 0.0073 |

Figure 3.3 shows the test-set performance gain for the different data splits. We can first notice that the optimal weight, selected based on the gain on the validation set, turned into a slight loss in the test set for only one split in the Instant Video dataset ($\Delta = -0.0054$), and one split in the Digital Music dataset ($\Delta = -0.001$). Detailed results not reported in the chapter show that this is mainly due to a drop in $\Delta_4$, representing the users that are hard to optimize for in any case. In 5 cases the optimal weight was $w = 0$, which yields a gain $\Delta = 0$.[5] For the majority of cases though (23 out of 30 splits), the optimal weight selection achieves a higher gain $\Delta$ and therefore helps achieve a better overall balance in recommendation performance across mainstream and non-mainstream users. If we consider only top 90% users to select the best weight, there are in total 28 out of 30 splits (except 2 in the Digital Music dataset) where NAECF shows superiority over DeepCoNN. However, this does not mean that a higher weight is always better for reaching a balance across users. For the BeerAdvocate dataset, 83% of the top-3 weights selected via the validation process are not larger than 0.5. For the two Amazon datasets, and although the optimal weights distribute over all weight candidates, unreported results still show that weights no larger than 2.0 take 85% of the top-3 best results. This observation matches our expectation that a mild weight value is more likely to bring a better trade-off between the overall recommendation accuracy and the balance across users. Based on these observations, we provide a positive answer to RQ1.

**EFFECT OF TEXT REVIEWS.**
We hypothesize that exploiting elaborate user- and item-related information, in our case in the form of online reviews, not only contributes to overall recommendation performance [146, 26], but also to neutralizing the mainstream bias. While non-mainstream users are relatively underrepresented in the user space, it should at least help if their individual representations are as elaborate as possible to model their preferences better.

---

[5] Due to the stochastic nature of the training process, one retraining of DeepCoNN may yield a slight gain with respect to another, but it should be zero on expectation. Therefore, we set $\Delta = 0$ when $w = 0$.

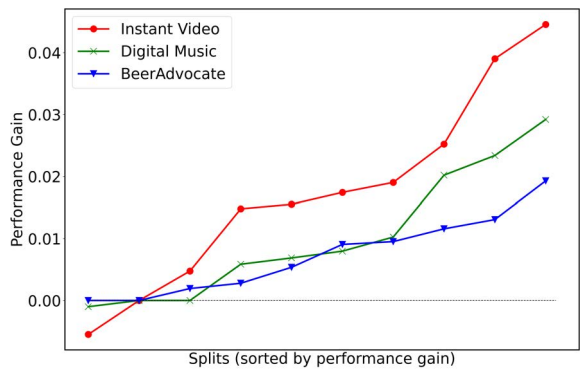

Figure 3.3: Test-set performance gain Δ on each of the 10 data splits, sorted within dataset.

In order to verify this hypothesis, we investigate the effect of this additional information compared to the case where it is not used, such as in a classical collaborative-filtering model like MF. We deliberately do not compare MF with NAECF because we would confound the use of text reviews for boosting recommendation accuracy and balancing across users. Instead, we choose to compare with DeepCoNN, which may anyway be regarded as a special case of NAECF and, architecture-wise it is the closest to collaborative filtering in the NAECF family. As such, a superiority of DeepCoNN over MF will indirectly mean a superiority of NAECF as well.

Fig. 3.4 shows the $uRMSE$ improvements on the test set made by DeepCoNN, compared to MF on all 3 datasets. As the figure shows, our expectations are met on all three datasets. The improvement on $uRMSE$ scores has a clearly positive correlation with the baseline $uRMSE$ achieved by MF, meaning that it is the users who received worse recommendations in MF, the ones who benefit the most from the inclusion of textual features in DeepCoNN. We note that, close to the origin of the plots, we see that DeepCoNN leads to slight performance loss for the users for which MF achieved the best performance. This is however an artifact of the evaluation process. Note that users with a $uRMSE$ close to zero in MF have almost no room for improvement, so any other model we compare with will probably perform worse. Similarly, other models will likely perform better for the users with very high $uRMSE$ in MF, because it is just not possible to perform worse. To illustrate and account for this effect, Fig. 3.4 also compares with a retrained MF model, displaying both the overall correlation and the loss close to the origin. These serve as a sort of baseline to assess the improvement of DeepCoNN (ie. rather than comparing the red curve with the $y = 0$ axis, compare it with the blue curve). We can thus confirm the superiority of DeepCoNN, as $uRMSE$ scores are always higher than on a retrained MF model.

Finally, and similar to the comparison between DeepCoNN and NAECF stated in Eq. (3.10), here we also compare DeepCoNN and MF in terms of gain Δ. The values on three datasets are 0.1175, 0.1017 and 0.3282, respectively. Such a significant improvement shows the effectiveness of review-based features in creating balance across different users, by which we provide an answer to RQ2.

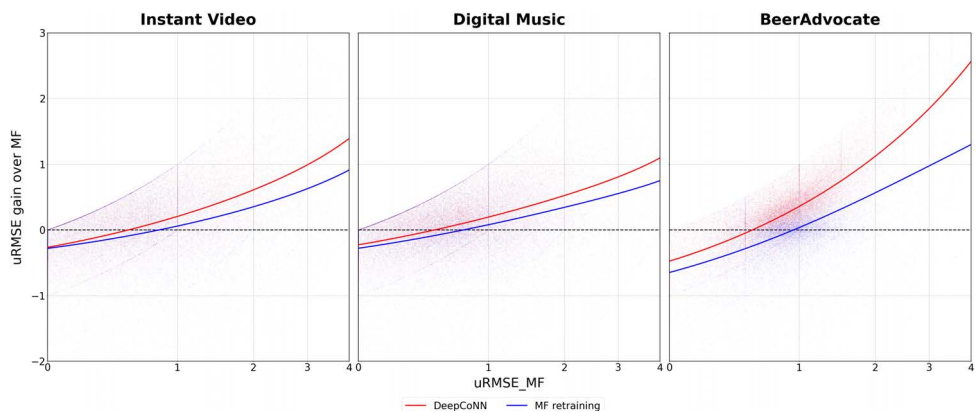

Figure 3.4: *uRMSE* gain over MF (positive is better) of DeepCoNN and a retrained MF model, on all 10 data splits. Curves represent a spline-smoothed fit.

In summary, we confirmed that NAECF creates a better balance across users by significantly improving the recommendation accuracy for non-mainstream users, subject to a good selection of the weight hyper-parameter. We also compared the review-based DeepCoNN and the CF-based MF, and found that the improvement stems mainly from a better optimization for non-mainstream users who are harder to handle in bare collaborative filtering. This way, NAECF's superiority lies in the use of review text, not only to boost rating prediction, but also as an adversary to ensure better user representation. Ultimately, these findings direct an open question to the correlation between the text reconstruction loss and the recommendation accuracy, which we study next.

### 3.5.2. USER FEATURE RECONSTRUCTION

Mainstream users are generally active and display common behavioral patterns. This makes it easier for them to be matched with proper neighbors in collaborative filtering, ultimately giving them more accurate recommendations. At the same time, good performance on similarity-based user modeling will make them easier to reconstruct in the NAECF autoencoders, and should therefore have a lower text reconstruction loss after training. This should be reflected by a positive correlation between *uRMSE* scores and user reconstruction losses $loss_U$. Because DeepCoNN does not contain any text reconstruction module, the loss should be randomly distributed and uncorrelated with *uRMSE*; we verified this in the data but do not report it here. However, intuition tells us that mainstream users should have low reconstruction losses. The failure of Deep-CoNN to reflect this expectation means there is room for improvement to create a better balance across users, which is confirmed by our findings in the previous section.

Therefore, we now look into the correlation between user reconstruction loss $loss_U$ and *uRMSE* recommendation accuracy. Fig. 3.5 shows the relationship for each of the evaluated weights $w$. We can see clear differences across datasets, but there are several qualitative commonalities. First, the majority of users are not mainstream even if they have a rather low *uRMSE* score. Second, higher weights generally lead to lower recon-struction losses and therefore to better user representations. This is expected because a high weight makes the text reconstruction losses dominate the overall loss in Eq. (3.1), but the figure further shows that the relative relationship between $loss_U$ and $uRMSE$ is pretty consistent across weights. Interestingly, the BeerAdvocate dataset shows some fluctuations with high weights. This evidences that the optimal weight needs proper tuning, because excessively high weights lead to substantial performance loss and $uRMSE$ scores become less stable as a consequence. As reported in Table 3.3, the optimal weights for this dataset are rather small indeed in comparison with the other two datasets.

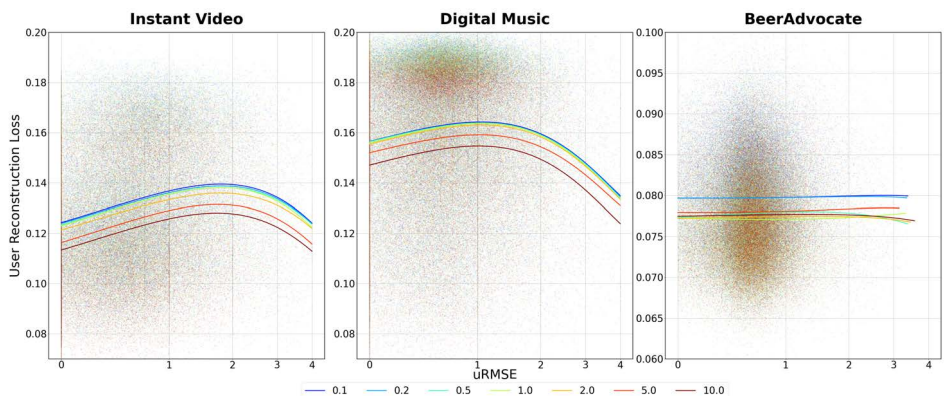

Figure 3.5: NAECF $loss_U$ by test-set $uRMSE$, for each weight $w$. Lines represent a spline-smoothed fit.

We also followed the earlier approach of dividing users in four bins according to the $uRMSE$ distribution. Fig. 3.6 similarly shows the relationship between $loss_U$ and $uRMSE$ with all the evaluated weights, but differentiating among user bins. We can clearly observe that, as expected, the relationship is monotonically positive except for the last bin in the Digital Music dataset. This confirms again that users who are better represented receive more accurate recommendations. As reported in Table 3.4, mainstream users in bin 1 do not always benefit from NAECF because they already receive good recommendations and there is little room for improvement, regardless of how well they are reconstructed. Fig. 3.6 confirms this especially in the two Amazon datasets, where bin 1 users receive nearly perfect recommendations. But NAECF improves performance especially for the 80% of non-mainstream users in bins 2 and 3, because those are harder to represent to begin with. Fig. 3.6 confirms that these users generally have the highest reconstruction losses indeed. Together with the correlations in Fig. 3.5, we see the relationship between the mainstreamness of users and the difficulty to represent them. Notwithstanding, the bottom 10% of users in bin 4 are too extreme to find proper representations, so the autoencoders hardly work for them. We even observed in Fig. 3.6 a negative correlation on the Instant Video and Digital Music datasets when $uRMSE$ is large. This confirms that NAECF sacrifices performance for these extreme users in favor of the others. Although unfortunate, we find this behavior acceptable because these users often display such particular tastes and patterns that it is hard for them to benefit from virtually any CF-based recommendation model.

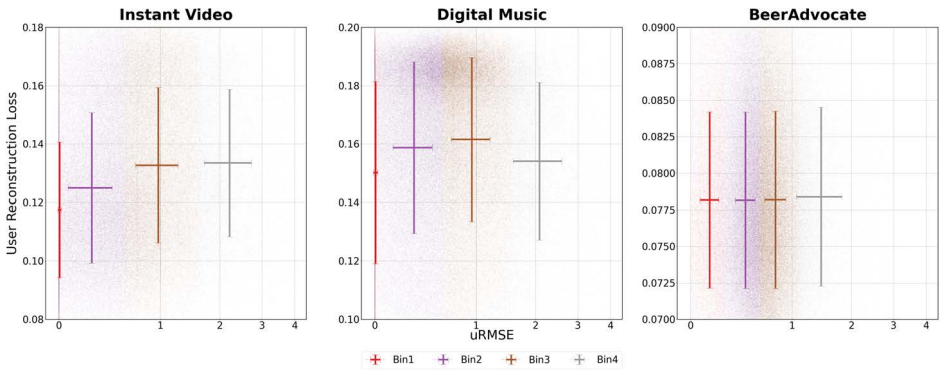

Figure 3.6: NAECF $loss_U$ by test-set $uRMSE$, with all evaluated weights $w$. Error bars show standard deviations per user bin.

## 3.6. CONCLUSION AND FUTURE WORK

Rating accuracy has long been an important criterion to evaluate recommender systems, if not the most important. Previous research has therefore focused mainly on maximizing the overall performance averaged over users. However, traditional collaborative filtering methods focus more strongly on recommending items that have positive interactions by similar users. In this situation, it is hard for CF models to work well with non-mainstream users that have special tastes or habits. Because non-mainstream users are rather a minority, this problem may not have a strong effect on the overall accuracy, yet it may create an unfair imbalance across users. To address this problem, we proposed a conceptually simple but effective model named NAECF, which minimizes the rating prediction loss while keeping the user and item properties preserved in the learned user and item representations. Preservation of user and item properties is imposed as an adversarial condition by minimizing reconstruction losses in addition to rating prediction error. This prevents these representations from being biased towards mainstream users.

We conducted experiments on three real-world datasets, and found that NAECF achieves an overall rating accuracy that is on par with the state-of-the-art. However, its strength is in the better balance it achieves across users thanks to a significant improvement of the recommendation accuracy for non-mainstream users, without significantly harming the mainstream ones. This improvement is achieved through an optimal trade-off between rating prediction and text reconstruction. Our results confirm a clear correlation between how well users are represented and the quality of their recommendations, evidencing that side information may be instrumental not only for boosting overall accuracy, but also to minimize possible biases in the learned models.

Future work will be conducted in several directions. First, we will investigate whether the conclusions drawn here for rating prediction generalize to the ranking paradigm, which is gaining popularity in the recommendation field. Second, in this chapter we treated users and items as equally important through a single text reconstruction weight. One may argue that improving the representation of users alone is not enough, because the model also needs a good item representation to know what to recommend. However, users and items may have different impacts, and we would like to explore this question by implementing two weights in the NAECF loss. Third, we introduced side information from text reviews in order to achieve a better balance across users. However, text reviews are just an example of additional content-based resources such as images and demographic information that can be used to achieve a similar function. We would like to further investigate the effect of other side information in the future and, perhaps more importantly, how to effectively incorporate such information in NAECF to eliminate the mainstream-bias. Finally, we are also interested in combining NAECF with explainable recommendation, so that we can provide convincing explanations to non-mainstream users.

# 4
# MITIGATING MAINSTREAM BIAS IN RECOMMENDATION VIA COST-SENSITIVE LEARNING

## 4.1. INTRODUCTION

One of the critical limitations of recommender systems based on collaborative filtering (CF) models [41] is that they are *not fair* in how they serve different groups of users [64, 69]. This fairness issue is a result of the varying quality of users' neighborhoods (groups of users with similar preferences) from which information is taken to train a CF model [150, 65]. The information collected from large, coherent, and information-rich neighborhoods will be the dominant one in steering the process of learning to recommend for all users. We refer to such dominant neighborhoods as *mainstream*. Because the users belonging to such neighborhoods —the *mainstream users*— are compatible with the learned model, they are optimally served. For the *non-mainstream* users, e.g. *niche* groups who deviate from the mainstream and whose interaction information is therefore less rich [69], who are less active compared to the mainstream users [90], or where the preferences are not well pronounced, the neighborhoods cannot fully reflect their genuine preferences. All this will make the non-mainstream users receive recommendations of a lower quality than the mainstream users. The difference in the quality of the CF model for these two user groups, further referred to as the *mainstream bias*, will result in the continuous improvement of the performance for the mainstream group, and continuous decrease of the performance for the rest [76].

While the issue of treating users differently by a recommender system in general has been addressed by a number of approaches, making for example assumptions about the relation between users' gender [83] or demographics [38] and the quality of recommendation, not many approaches have focused specifically on addressing the mainstream bias. Li, Urbano, and Hanjalic [65] deployed an autoencoder [103] for feature reconstruction as an adversary to a traditional CF model, forcing it to deviate from the pure similarity-based learning and make the learned model more compatible with the non-mainstream users. More specifically, the autoencoder was deployed to steer the process of learning the user/item representation space for rating prediction via optimal reconstruction of the properties of all users, mainstream and otherwise, assuming this would lead to equal treatment of users during recommendation. Still, a more explicit focus on the mainstreamness of users is needed to ensure that the bias is effectively addressed.

Inspired by outlier detection techniques, Zhu and Caverlee [150] did focus on explicitly quantifying mainstreamness via similarities of user-preference profiles, and incorporated them to fine-tune the recommendation process for different user groups. However, in the absence of ground truth data about mainstreamness, it is difficult to assess how well these approaches identify non-mainstream users. In addition, these mainstreamness statistics are model-agnostic in the sense that they are independent of the recommendation strategy, effectively ignoring the model's own capability to reduce the mainstream bias or even amplify it. As a result, the learning process could be tailored to the wrong users.

In this chapter, we choose to focus there where the effect of mainstreamness is *directly* observed, that is, the recommendation utility provided by the data and recommendation model at hand. If a user receives poor recommendations it could be because their preferences deviate from the rest, or because there is not enough data to properly quantify their similarity to other users or to fully exploit it. Therefore, we choose utility as an implicit proxy for mainstreamness. Through this quantification of user mainstream-ness, we make the training process focus on the non-mainstream ones by assigning them higher weights. We do so, however, in a *cost-sensitive* way [123], taking the cost of recommendation errors into account while training the CF model. Our results show that our implicit measurement of mainstreamness via utility is better able to differentiate niche users than an explicit approach, and that the cost-sensitive learning strategy does mitigate the bias by balancing the recommendation quality across users. Finally, we investigate data requirements for conducting research on mainstream bias at the individual user level, and provide suggestions for reliable experimentation in this area.

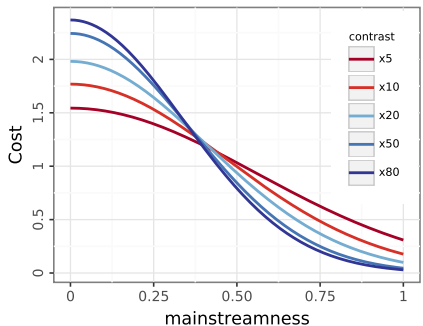

Figure 4.1: Cost functions used in the paper. The contrast denotes the relative cost between users with mainstreamness 0 and users with mainstreamness 1 (i.e. x10 means 10 times as much).

## 4.2. PROPOSED APPROACH

The basis of our approach is a weighted loss function where every user $u \in U$ is assigned a weight $\omega(u)$ that informs the learning process about the importance of every user's individual recommendation loss. The global loss is thus simply

$$L = \sum_{u \in U} \omega(u) L_R(u) \,, \tag{4.1}$$

where the recommendation loss $L_R$ is specific of the model and learning paradigm. This way, we explicitly tell the learning process what users to optimize for by means of $\omega$, which, in our case, should be high for non-mainstream users and low for mainstream users.

### 4.2.1. DEFINITION OF WEIGHTS

As explained in the previous section, we define $\omega$ as a function of the user mainstreamness $m_u$. However, rather than simply using a naïve transformation of $m_u$, we introduce flexibility through a cost function that maps user mainstreamness onto a cost value. In particular, and assuming $m_u$ ranges between 0 and 1, we use the density function of a Normal distribution truncated between 0 and 1, with zero mean and variance adjusted to achieve a contrast ranging between 5 (i.e. users with mainstreamness $m_u = 0$ have a cost 5 times as large as users with $m_u = 1$) and 80 (ie. 80 times as much). This is a simple choice to make $\omega$ smooth and monotonically decreasing, but other cost functions that emphasize different levels of mainstreamness are of course possible; we leave this discussion for further work. Fig. 4.1 shows some examples. Nonetheless, the formulation of the cost function may consider various aspects tailored to the business case, as well as different magnitudes for the contrast between users with low and high mainstreamness. For example, it would be reasonable to assign very high weights to non-mainstream users with high activity, or to users with very low activity as an attempt to reduce the churn rate.

An important point to consider when defining $\omega$ is the distribution of mainstreamness across users. It could be the case that, given the current data and model, the least mainstream users are actually fairly mainstream already, so their weight relative to the most mainstream users should be adjusted via a smaller contrast. It could also be the case that the dataset is very sparse and there are simply not enough neighbors around users for the model to learn a good representation. That is, the majority of users could be considered non-mainstream, and as a result the cost function would hardly differentiate among them. Lastly, one could decide to compute $m_u$ in several different ways (see next Section), which could potentially lead to quite different mainstreamness score distributions altogether, ultimately leading to a different set of weight values even for the same users.

In order to minimize this dependence on the dataset and mainstreamness definition, and ensure that the full co-domain of the cost function is used, we first normalize the raw mainstreamness scores. Simply re-scaling between the minimum and maximum could still lead to a disproportionate use of small parts of the co-domain, and would also be very sensitive to outlier users. Instead, we use the rank statistic of $m_u$ normalized in $[0,1]$. We achieve this by using the empirical cumulative distribution function (ecdf)

$$\omega(u) = \text{cost}(\text{ecdf}_{\mathcal{U}}(m_u)), \qquad (4.2)$$

where, as mentioned, cost is defined in terms of a truncated Normal density function.

### 4.2.2. MEASUREMENT OF MAINSTREAMNESS

An **explicit** approach to compute $m_u$ would ideally follow some notion of mainstreamness, but mainstreamness is itself a complex construct very hard to define formally [14, 150, 65]. Recently, Zhu and Caverlee [150] took inspiration from outlier detection techniques to propose four different definitions:

- Sim: users are mainstream to the extent that their interactions are similar to that of the other users. The Jaccard coefficient is used to measure the average similarity between a user and all the others.

- Den: users are mainstream to the extent that there are enough close neighbors to calculate similarity with. The local outlier factor algorithm (LOF) [18] is used to identify niche users.

- Dis: users are mainstream to the extent that their interactions are common in the dataset, that is, they interact with popular items. The cosine similarity is used to measure the similarity between a user and the average user interactions.

---

[1]Data available from the authors' public repository at https://github.com/Zziwei/Measuring-Mitigating-Mainstream-Bias.

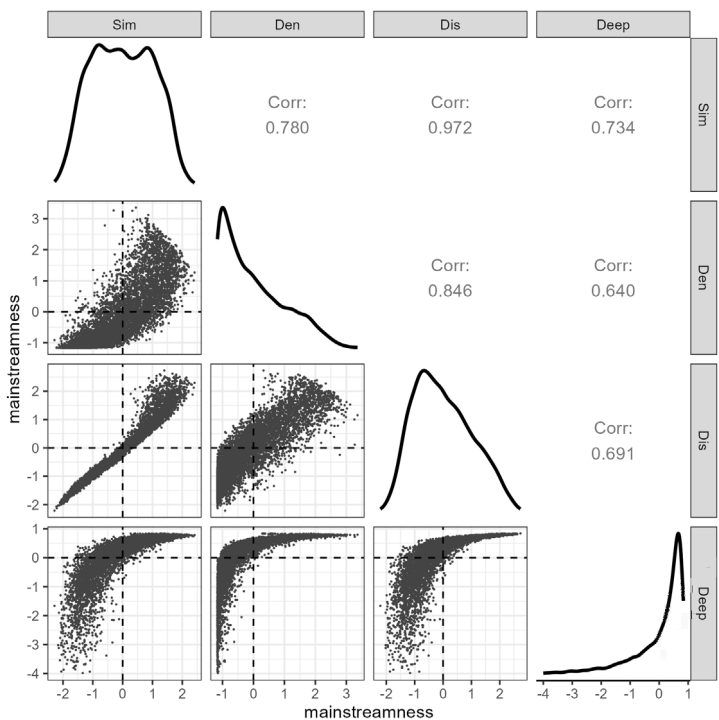

Figure 4.2: Comparison of the four mainstreamness definitions proposed by Zhu and Caverlee [150], as applied to the MovieLens 1M dataset.[1] Density plots illustrate the distribution of mainstreamness for each definition. Scatter plots show the relationship between pairs of definitions, quantified in the upper-right half via Pearson correlation scores. Scores are standardized to zero mean and unit variance for better comparison.

- Deep: similar to Den, niche users are identified by an outlier detection algorithm. In particular, the deep support vector data description algorithm (DeepSVDD) [102] is used.

However, it is difficult to assess how well these, or any other definitions for that matter, correlate with the concept of mainstreamness. To illustrate, Fig. 4.2 compares these four definitions as applied to the MovieLens 1M dataset. Although they are somewhat correlated to one another, it is evident that they produce very different scores. For instance, Sim and Dis lead to nicely shaped distributions, suggesting few users with extreme (non-)mainstreamness. However, Den and Deep lead to very skewed distributions, even in the opposite direction, pointing to many users with extreme scores. This shows that the same user could be considered both mainstream or non-mainstream, depending on how we choose to define mainstreamness.

Furthermore, it should be noted that these four definitions of mainstreamness are agnostic to the recommendation model. However, the effect of mainstreamness, ultimately, depends on the model and how it is able to exploit the specifics of the dataset it is trained on. It is not far-fetched to think of a user, assessed as non-mainstream, who receives bad recommendations under one model but good recommendations under a more capable one.

This leads us to consider an alternative, **implicit** way to quantify mainstreamness that is *not* model agnostic. In particular, we decide to focus there where the effect of mainstreamness is to be observed, that is, the recommendation *utility* provided by the recommendation model at hand. This is where mainstreamness will ultimately have an impact on. The very nature of collaborative filtering tells us that if a user receives poor recommendations it is because they are non-mainstream under the current model: they cannot be properly represented, either because their preferences are somehow different from their closest neighbors, or because there are not enough data to properly quantify their similarity. Therefore, we use utility as a proxy for mainstreamness. Since utility, just like mainstreamness, is a complex concept difficult to measure, we decide to simply use the accuracy of the recommendation model for that user, measured through a metric like *nDCG* or *AP*.

But there is the question of what accuracy scores we actually use. In principle, these scores should reflect user mainstreamness when there is no mechanism to minimize its effect, and they should be achieved by the recommendation model in the dataset at hand. Therefore, we decide to use the accuracy achieved, *on a validation set*, by the vanilla model whose loss function is as in Eq. (4.1) but using no weights. As intended, we thus first see how the model reacts to mainstreamness as reflected in the observed utility for users, and then act upon it in a cost-sensitive way.

## 4.3. EXPERIMENTAL DESIGN

We carried out a number of experiments to investigate the effectiveness of the proposed approach in mitigating the mainstreamness bias, as well as the effect of the contrast applied by the cost function. In particular, we study contrasts x5, x10, x20, x50 and x80, that is, the most non-mainstream user has a weight between 5 and 80 times larger than that of the most mainstream user. Fig. 4.1 details the cost functions. Regarding the measurement of mainstreamness, we consider both an explicit and an implicit quantification. For the former, we follow Zhu and Caverlee [150] and compute Sim scores. This choice is motivated by the time complexity of their four approaches (the computation of mainstreamness may quickly become intractable as the numbers of users and items increase; while their datasets include a few thousand items, ours span from a few thousands to over half a million), and their correlation to one another (Sim is also the one most correlated with the others, in particular with Deep). For the implicit quantification we compute utility scores using the metric *nDCG* as an exemplar of recommender systems research; hereafter, we will refer to this definition of mainstreamness as Util.

We selected four real-world datasets containing user-item rating interactions from various domains and with different densities, especially including some highly sparse datasets (see Table 4.1). In line with common practice in ranking-oriented recommender systems research, we see all existing interactions in the datasets as relevant, and all other unseen interactions as irrelevant. We use LensKit [37] to evenly split the relevant items for each user into training, validation and test sets. To make the modeling of utility — and hence mainstreamness— robust, each user has at least five relevant transactions in each of the three sets; we explain the rationale for this decision in Section 4.5. For training the model, we follow He et al. [47] and Wu, Wu, and Huang [139] and randomly sample four irrelevant items per relevant item in the training partition. For validation and test, we follow DaisyRec [119] and evaluate the model for each user by ranking a total of 500 items consisting of their relevant items in the validation/test partition and a set of randomly sampled irrelevant items. Finally, to make sure relevant items are the minority, as happens in reality, we truncate the number of relevant interactions to 200. The dataset statistics after processing are shown in Table 4.1.

Table 4.1: Dataset statistics after pre-filtering.

| Dataset | #users | #items | #ratings | Density |
|---|---|---|---|---|
| MovieLens 1M [46] | 6,040 | 3,609 | 562,957 | 2.583% |
| BeerAdvocate [79] | 8,821 | 43,663 | 780,752 | 0.203% |
| Amazon Digital Music [93] | 14,057 | 379,171 | 619,673 | 0.011% |
| Amazon Musical Instruments [93] | 15,270 | 585,766 | 862,798 | 0.010% |

Regarding the recommendation model, we deploy a simple but effective CF model that only utilizes user-item interactions. Specifically, we choose Factorization Machines (FM) [99], which optimize the binary cross-entropy (BCE) loss via the Adaptive Moment Estimation (Adam) [57] learner, and leave the investigation on other training paradigms for future work. For each user, the BCE loss is normalized by dividing by the total number of relevant and irrelevant items used for training, so that all user losses are on the same scale in (4.1). After a fine-tuning process based on grid search, we fixed several key hyper-parameters including the dimension of vectors used for interaction (32), learning rate (0.0001), L2-regularization coefficient to avoid overfitting (0.001), and batch size (512).

All models are trained for 300 epochs to ensure full convergence, and with 3 different random initializations to minimize random effects due to the sampling process. The whole pipeline is implemented in PyTorch [96], and all experiments are run on one NVIDIA GeForce GTX 2080Ti GPU [2].

## 4.4. RESULTS

### 4.4.1. MAINSTREAMNESS AND UTILITY

We first examine how Sim and Util differentiate between mainstream and non-mainstream users. In particular, we are interested in how well they correlate with the test $nDCG$ scores obtained by the baseline FM model: non-mainstream users should receive recommendations with low $nDCG$ scores, while mainstream users should receive higher scores.

For each of the four datasets, Fig. 4.3 compares Sim and Util. We can first see that both approaches lead to similar distributions in the Amazon datasets, where there appear to be many non-mainstream users. However, they somewhat disagree in the Beer-Advocate dataset, where Util does not identify many non-mainstream users to benefit from the cost-sensitive approach. In terms of correlation with the test $nDCG$ scores, we can see that Util is much better correlated, specially in the Amazon datasets. This points to the possibility that Sim identifies many non-mainstream users to which the model is still able to offer good recommendations. If the training process increases their importance by assigning them a high weight $\omega$, we may loose the opportunity to focus on those users that still receive poor recommendations.

---

[2] All data, code and results are available at
https://github.com/roger-zhe-li/ictir23-cost-sensitive.

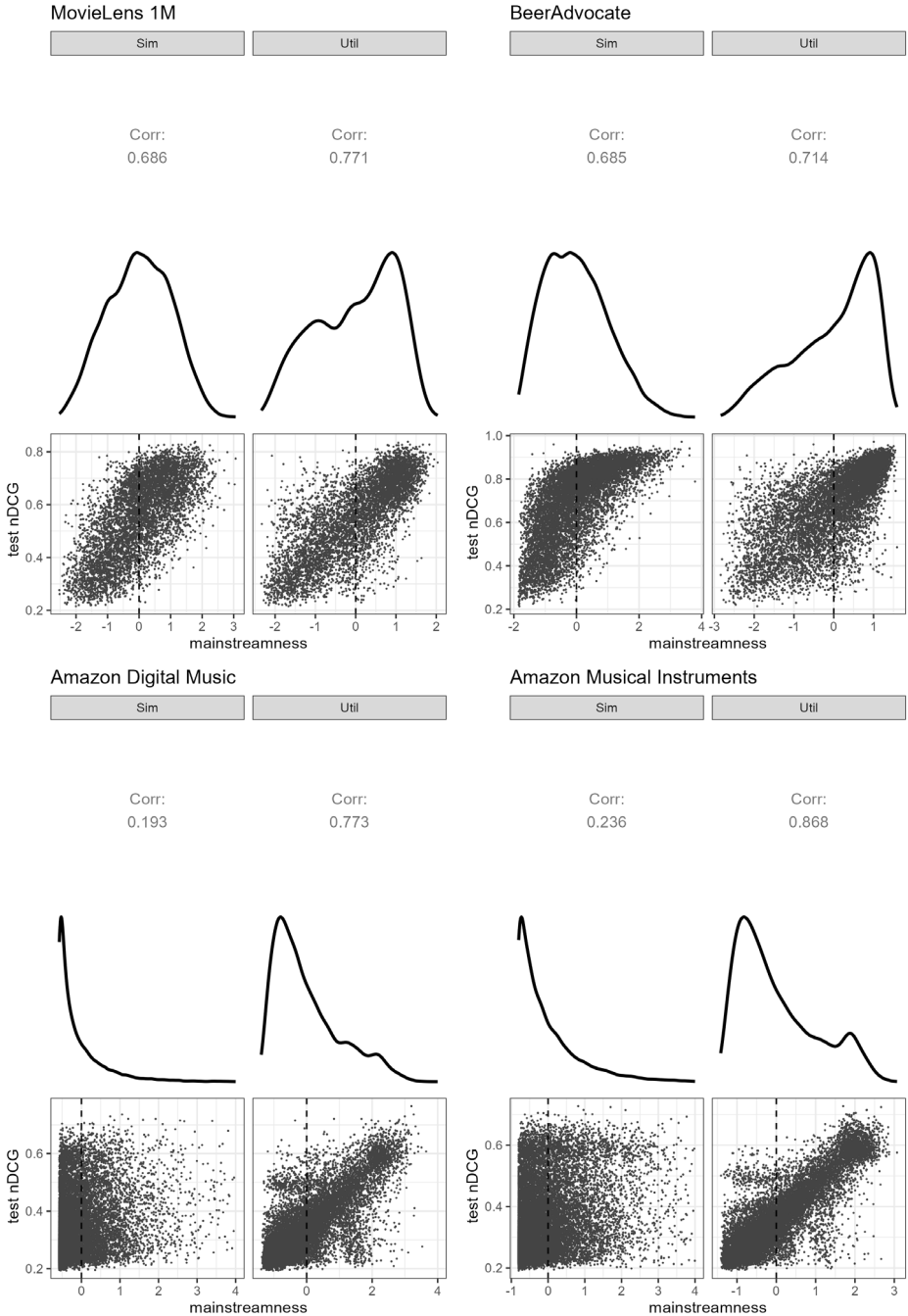

Figure 4.3: Correlation between mainstreamness and test nDCG in the baseline model (FM), for each mainstreamness definition. Density plots illustrate the distribution of mainstreamness. Scatterplots show their relationship with nDCG, quantified at the top via Pearson correlation scores. Mainstreamness scores are standardized to zero mean and unit variance for better comparison.

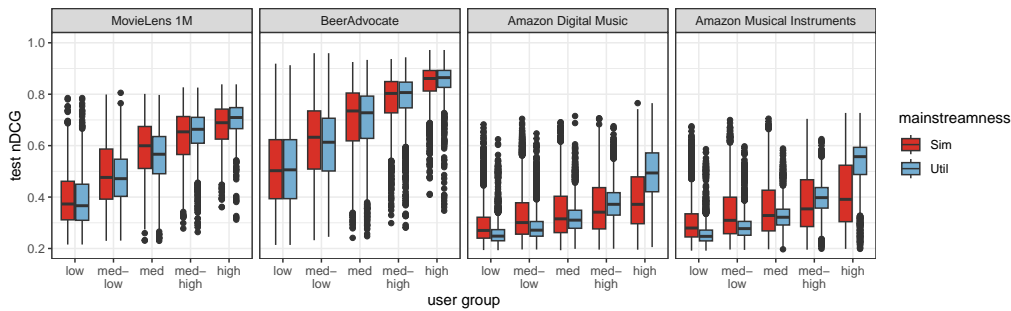

Figure 4.4: Correlation between user groups, split by mainstreamness, and test nDCG in the baseline model (FM).

In order to assess the effectiveness of the cost-sensitive approach for the mitigation of the mainstream bias, we will look in the next Section into different groups of users separated by their mainstreamness: group 'low' contains the 20% of users with lowest mainstreamness scores on the baseline model, group 'med-low' contains the next 20% or users, group 'med' contains the middle 20% of users, and so on with groups 'med-high' and 'high'. An effective mitigation of the mainstream bias would be reflected in increased performance for the lower groups, which ideally should be those with lowest test $nDCG$ scores in the baseline model. Fig. 4.4 shows how well Sim and Util separate users in these five groups. We can first see that the groups are indeed correlated with $nDCG$, but we can notice that this correlation is stronger with Util, specially in the Amazon datasets (the low groups receive lower utility, and the higher groups receive higher utility). We can also see that groups tend to overlap substantially when separated by Sim, potentially misplacing users. This overlap can be quantified by an ANOVA model of $nDCG$ modeled by two factors: dataset and user-group nested within dataset. Indeed, the user-group effect has a much larger sum of squares (SS) with Util than with Sim (SS=440 vs SS=218; SS of the dataset effect is 843). Finally, Fig. 4.4 also points that the BeerAdvocate dataset may be hard to further optimize for because the utility scores are already relatively high.

### 4.4.2. BIAS MITIGATION

An effective mitigation of the mainstream bias would be reflected in increased performance for the lower groups (i.e. mainly 'low' and 'med-low'), ideally with no detriment to the higher groups and, especially, overall. In the previous section we separated users into groups by each of Sim and Util, but here we separate them directly by their test $nDCG$ with the baseline model FM, because this better illustrates how non-mainstream users suffer from the bias.

Table 4.2 reports the relative percentage improvement in $nDCG$ scores per user group, as well as the overall mean score across all users in the dataset. We can clearly see that the use of Sim benefits the non-mainstream users only in the two Amazon dataset; in MovieLens and BeerAdvocate they are even hurt further. In contrast, Util is always able to improve the utility of non-mainstream users across datasets, achieving relative *nDCG* improvements of up to 5% in the Amazon datasets. Improvements on the lower user groups are generally higher than losses on the higher groups, where users already receive (very) high recommendation utility anyway and such minor losses are probably unnoticed. This redistribution of model performance has a negligible effect on the global performance of the models, as evidenced by the overall *nDCG* scores. This means that, with proper selection of the contrast in the cost function, Util can minimize the mainstream bias at virtually no overall loss in utility. On the other hand, the use of Sim for training leads to inferior overall performance on all four datasets.

Table 4.2: Mean nDCG of the baseline model (FM) per user group, and relative percentage improvement of each cost-sensitive model (e.g. users in group 'low' of MovieLens 1M received a score of .3284 with the baseline, and an improvement of +3.89% with the x80-contrast cost-sensitive model under the Util mainstreamness definition). Column 'Overall' lists the mean across all users. Green/red for statistically significant gain/loss with respect to the baseline (hierarchical linear model with seed and user random effects, Bonferroni correction).

|  |  | MovieLens 1M | | | | | BeerAdvocate | | | | |
|---|---|---|---|---|---|---|---|---|---|---|---|
|  |  | Overall | low | med-low | med | med-high | high | Overall | low | med-low | med | med-high | high |
|  | FM | .5531 | .3284 | .4621 | .5753 | .6613 | .7388 | .6887 | .4144 | .6051 | .7301 | .8132 | .8809 |
| Sim | x5 | .5465 | -0.36 | -0.89 | -1.62 | -1.3 | -1.33 | .6792 | -0.52 | -1.72 | -1.83 | -1.44 | -1.13 |
| | x10 | .5437 | -0.47 | -1.32 | -2.23 | -1.87 | -1.94 | .6734 | -0.81 | -2.87 | -2.94 | -2.27 | -1.8 |
| | x20 | .541 | -0.53 | -1.76 | -2.85 | -2.41 | -2.51 | .6666 | -1.35 | -4.23 | -4.13 | -3.22 | -2.56 |
| | x50 | .5376 | -0.67 | -2.28 | -3.64 | -3.06 | -3.26 | .6588 | -2.01 | -5.65 | -5.56 | -4.29 | -3.54 |
| | x80 | .5359 | -0.67 | -2.55 | -4.06 | -3.39 | -3.59 | .6548 | -2.55 | -6.39 | -6.17 | -4.84 | -4.08 |
| Util | x5 | .5567 | +1.67 | +1.81 | +0.63 | +0.16 | -0.13 | .6846 | +0.63 | -0.41 | -0.94 | -0.83 | -0.77 |
| | x10 | .5574 | +2.38 | +2.34 | +0.7 | +0.11 | -0.27 | .6807 | +0.44 | -1.19 | -1.66 | -1.39 | -1.27 |
| | x20 | .5579 | +3.05 | +2.87 | +0.73 | 0 | -0.48 | .6762 | +0.54 | -2.11 | -2.67 | -2.09 | -1.74 |
| | x50 | .5579 | +3.62 | +3.31 | +0.68 | -0.21 | -0.84 | .6722 | +2 | -3.09 | -3.86 | -2.89 | -2.32 |
| | x80 | .5577 | +3.89 | +3.47 | +0.63 | -0.32 | -1.05 | .6715 | +2.98 | -3.2 | -4.25 | -3.14 | -2.54 |
|  |  | Amazon Digital Music | | | | | Amazon Musical Instruments | | | | |
|  |  | Overall | low | med-low | med | med-high | high | Overall | low | med-low | med | med-high | high |
|  | FM | .3456 | .2324 | .2695 | .3145 | .3828 | .5289 | .3606 | .2348 | .2772 | .3276 | .4085 | .5552 |
| Sim | x5 | .3395 | +0.65 | +0.05 | -0.61 | -1.8 | -4.45 | .3581 | +1.61 | +0.77 | -0.09 | -1.06 | -2.5 |
| | x10 | .3368 | +0.99 | +0.1 | -1 | -2.67 | -6.27 | .3577 | +3.29 | +1.44 | +0.08 | -1.68 | -3.54 |
| | x20 | .3347 | +1.31 | +0.09 | -1.3 | -3.46 | -7.69 | .3576 | +3.33 | +1.44 | +0.04 | -1.72 | -3.58 |
| | x50 | .3328 | +1.63 | +0.07 | -1.6 | -4.13 | -8.92 | .3576 | +3.37 | +1.45 | +0.03 | -1.75 | -3.6 |
| | x80 | .3316 | +3.66 | +0.9 | -1.78 | -5.07 | -10.62 | .3576 | +3.39 | +1.45 | +0.02 | -1.77 | -3.61 |
| Util | x5 | .3454 | +1.59 | +1.43 | +1.2 | +0.4 | -2.58 | .3607 | +1.1 | +0.96 | +0.62 | -0.04 | -1.19 |
| | x10 | .3453 | +2.54 | +2.09 | +1.53 | +0.18 | -3.5 | .3607 | +2 | +1.52 | +0.78 | -0.32 | -1.8 |
| | x20 | .3454 | +3.59 | +2.78 | +1.64 | +0.11 | -4.25 | .3607 | +2.63 | +1.93 | +0.8 | -0.53 | -2.08 |
| | x50 | .3458 | +4.84 | +3.67 | +1.92 | -0.11 | -4.9 | .3608 | +3.94 | +2.48 | +0.85 | -0.83 | -2.66 |
| | x80 | .346 | +5.45 | +4 | +2.02 | -0.18 | -5.15 | .3608 | +4.48 | +2.64 | +0.88 | -0.97 | -2.86 |

Fig. 4.5 presents a more fine-grained picture with one of the three random initial-izations in our experiments. Curve segments above 0 represent an improvement by the cost-sensitive models, while segments below 0 represent a loss. We can confirm that the cost-sensitive approach indeed makes the models focus on the non-mainstream users, as shown by the nicely smooth correlation between observed utility and relative improvement, moderated by the contrast in the cost function. As expected though, this focus on the non-mainstream users comes at the cost of a utility loss for the mainstream users on the right-hand side of the plots. Nevertheless, when using Util the relative loss for those users is generally much smaller than the gain for the very non-mainstream users, which are our main target. The figure also shows that the actual relation between improvement and utility varies across datasets, as reflected by the different curve shapes. This is explained by the differences in the shape of their $nDCG$ distributions (see Fig. 4.3); recall that we use the $ecdf$ of the scores. In a side-by-side comparison between Sim and Util, we see that Util offers better performance nearly everywhere along the $x$-axis, but especially for the non-mainstream users.

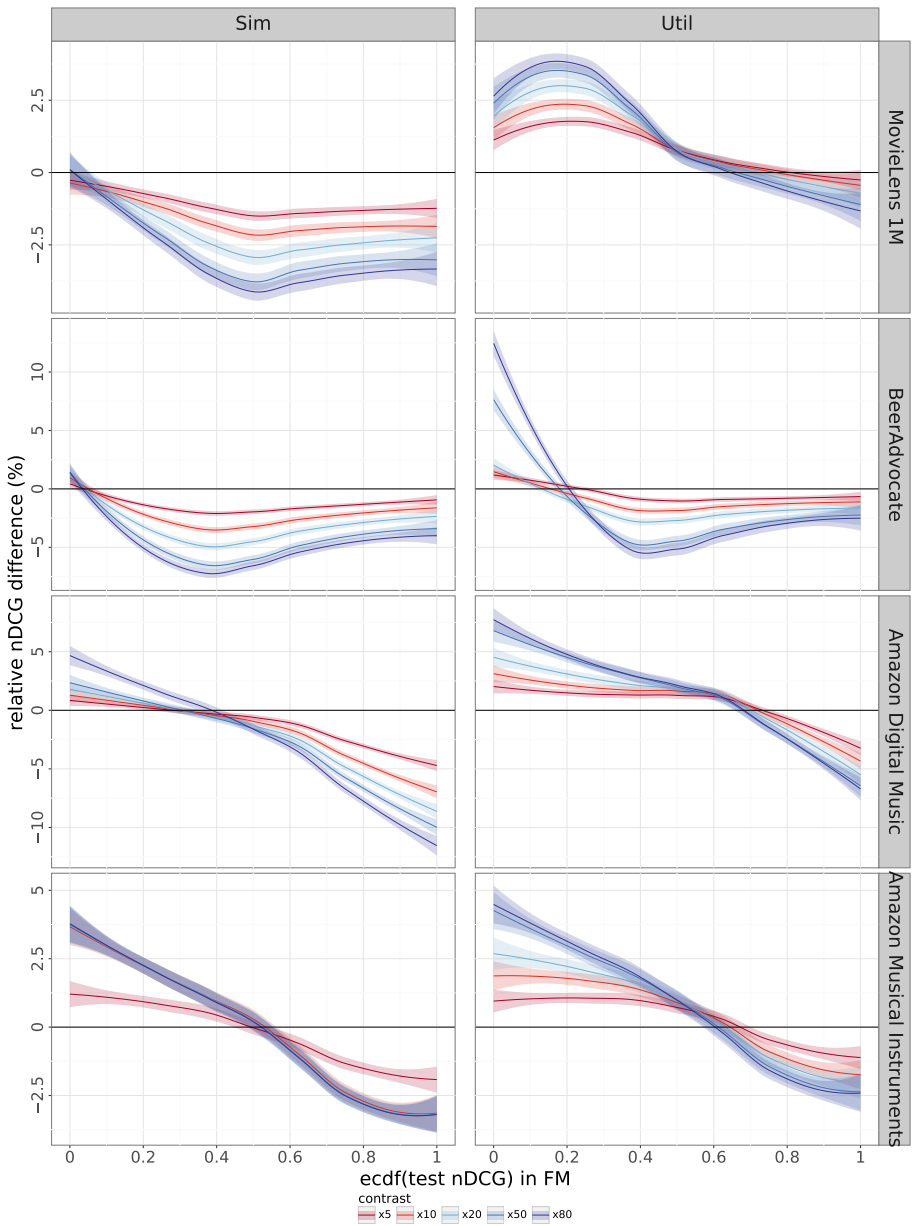

Figure 4.5: Mean nDCG relative percentage improvement between cost-sensitive models and baseline model, as a function ecdf(test nDCG) in the baseline FM model, for a sample data split. Curves fitted by a LOESS model. Ribbons indicate 95% confidence intervals.

Table 4.3: Same as Table 4.2, but user groups defined by Sim scores instead of test nDCG in the baseline model.

|  |  | MovieLens 1M | | | | | BeerAdvocate | | | | |
|---|---|---|---|---|---|---|---|---|---|---|---|
|  |  | Overall | low | med-low | med | med-high | high | Overall | low | med-low | med | med-high | high |
|  | FM | .5531 | .3284 | .4621 | .5753 | .6613 | .7388 | .6887 | .4144 | .6051 | .7301 | .8132 | .8809 |
| Sim | x5 | .5465 | -0.62 | -0.97 | -1.16 | -1.34 | -1.58 | .6792 | -1.4 | -1.39 | -1.43 | -1.34 | -1.35 |
| Sim | x10 | .5437 | -0.84 | -1.42 | -1.63 | -1.87 | -2.31 | .6734 | -2.23 | -2.4 | -2.23 | -2.14 | -2.16 |
| Sim | x20 | .541 | -1.19 | -1.83 | -2.12 | -2.35 | -2.96 | .6666 | -3.42 | -3.54 | -3.17 | -2.95 | -3.06 |
| Sim | x50 | .5376 | -1.49 | -2.35 | -2.71 | -3.02 | -3.84 | .6588 | -4.71 | -4.74 | -4.31 | -3.92 | -4.2 |
| Sim | x80 | .5359 | -1.76 | -2.68 | -2.95 | -3.32 | -4.19 | .6548 | -5.34 | -5.29 | -4.83 | -4.46 | -4.91 |
| Util | x5 | .5567 | +1.5 | +0.99 | +0.53 | +0.32 | +0.26 | .6846 | -0.46 | -0.5 | -0.51 | -0.66 | -0.72 |
| Util | x10 | .5574 | +2.05 | +1.24 | +0.67 | +0.33 | +0.2 | .6807 | -1.11 | -1.16 | -1.1 | -1.19 | -1.22 |
| Util | x20 | .5579 | +2.52 | +1.59 | +0.71 | +0.27 | +0.08 | .6762 | -1.85 | -2.03 | -1.79 | -1.73 | -1.72 |
| Util | x50 | .5579 | +2.92 | +1.8 | +0.64 | +0.12 | -0.16 | .6722 | -2.66 | -2.85 | -2.53 | -2.13 | -2.05 |
| Util | x80 | .5577 | +3.12 | +1.86 | +0.61 | +0.01 | -0.33 | .6715 | -2.92 | -3 | -2.65 | -2.15 | -2.05 |
|  |  | Amazon Digital Music | | | | | Amazon Musical Instruments | | | | | |
|  |  | Overall | low | med-low | med | med-high | high | Overall | low | med-low | med | med-high | high |
|  | FM | .3456 | .2324 | .2695 | .3145 | .3828 | .5289 | .3606 | .2348 | .2772 | .3276 | .4085 | .5552 |
| Sim | x5 | .3395 | -0.04 | -0.61 | -1.37 | -2.16 | -4.05 | .3581 | +0.01 | -0.16 | -0.52 | -0.89 | -1.65 |
| Sim | x10 | .3368 | 0 | -0.86 | -2.03 | -3.18 | -5.71 | .3577 | +0.26 | -0.08 | -0.53 | -1.14 | -2.11 |
| Sim | x20 | .3347 | -0.05 | -1.15 | -2.52 | -4.01 | -6.97 | .3576 | +0.24 | -0.11 | -0.57 | -1.16 | -2.14 |
| Sim | x50 | .3328 | -0.17 | -1.42 | -3.01 | -4.67 | -8 | .3576 | +0.24 | -0.09 | -0.58 | -1.18 | -2.16 |
| Sim | x80 | .3316 | +0.03 | -1.52 | -3.34 | -5.18 | -8.87 | .3576 | +0.23 | -0.1 | -0.59 | -1.18 | -2.17 |
| Util | x5 | 0.3454 | +0.12 | +0.11 | -0.16 | +0.07 | -0.3 | 0.3607 | -0.01 | +0.06 | 0 | -0.04 | +0.12 |
| Util | x10 | 0.3453 | +0.13 | +0.06 | -0.11 | +0 | -0.43 | 0.3607 | 0 | +0.12 | +0.01 | -0.06 | +0 |
| Util | x20 | 0.3454 | +0.11 | +0.06 | -0.06 | +0.15 | -0.5 | 0.3607 | 0 | +0.14 | +0.05 | -0.08 | +0.01 |
| Util | x50 | 0.3458 | +0.19 | +0.07 | +0.13 | +0.22 | -0.3 | 0.3608 | +0.16 | +0.23 | -0.03 | -0.05 | -0.04 |
| Util | x80 | 0.346 | +0.23 | +0.09 | +0.15 | +0.29 | -0.18 | 0.3608 | +0.16 | +0.21 | +0.01 | -0.02 | -0.06 |

In summary, we see that our cost-sensitive approach brings better balance across users, thus helping in the mitigation of the mainstream bias. In addition, we confirm that an implicit quantification of mainstreamness like Util works better than an explicit quantification like Sim in steering the learning process towards better recommendations for the users that receive low utility from the baseline model. In addition, we note that the mitigation effect via Util does not decay with increasing data sparsity (refer back to Table 4.1).

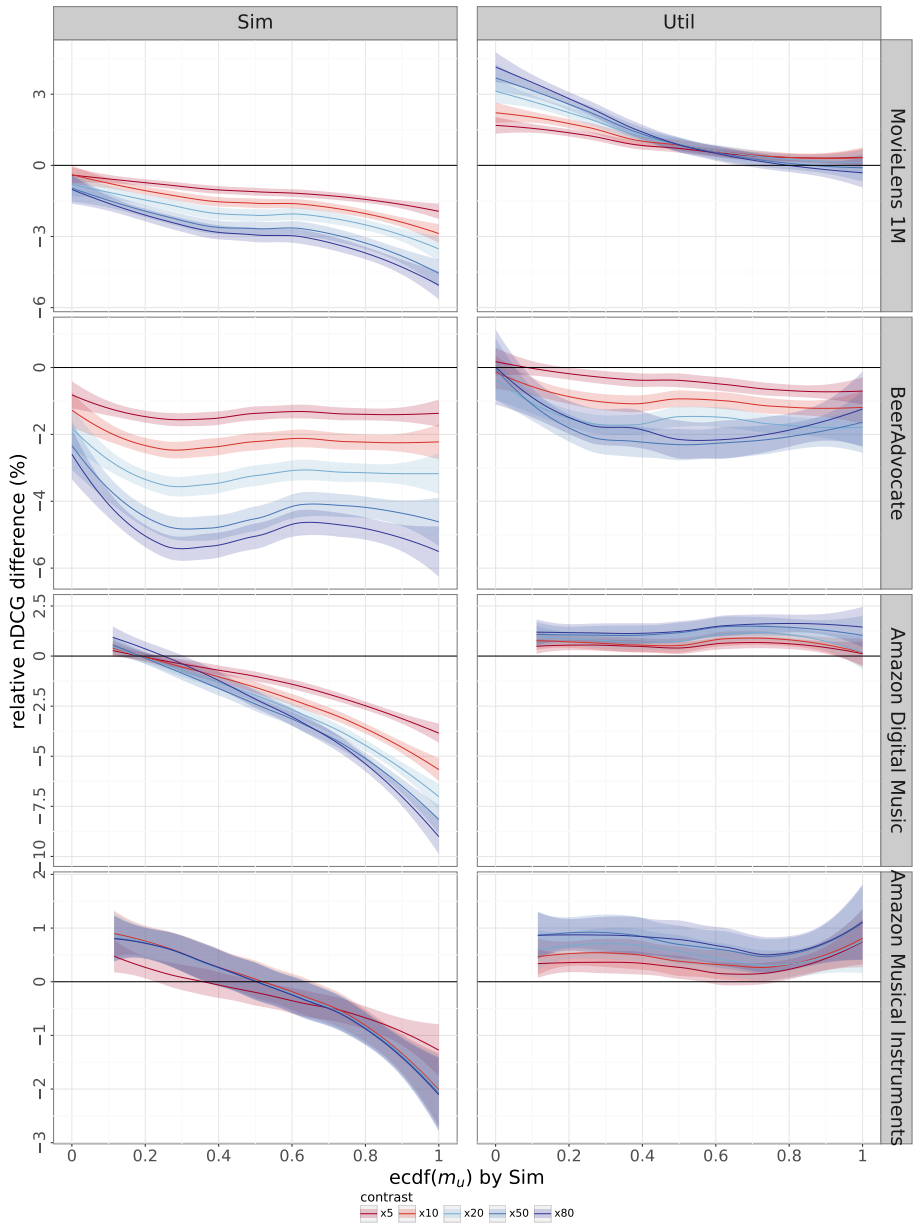

Figure 4.6: Same as Fig. 4.5, but plotted against ecdf($m_u$) by Sim instead of ecdf(test nDCG) in the baseline model.

One could be tempted to argue that Util should obviously offer better results than Sim when analyzing *test nDCG* because it is based on *validation nDCG* scores; test and validation scores should be highly correlated (we will come back to this in Section 4.5). After all, both Table 4.2 and Fig. 4.5 analyze results by test *nDCG*. The argument made above is that differences between mainstream and non-mainstream users can be immediately identified by test scores, but for the sake of clarity and to avoid potentially unfair assessment towards Sim, Table 4.3 reports the same results but separating users by Sim, while Fig. 4.6 does so by plotting against Sim. While the results are less clear with this partition of users, the table confirms that models trained with Sim are generally better at mitigating the bias than those trained with Util. In particular, results for the BeerAdvocate dataset show that higher contrasts even lead to worse performance for the lower user groups, suggesting that Sim is perhaps not properly identifying non-mainstream users. The figure shows that Util improves over the baseline across all levels of mainstreamness in the Amazon datasets, further suggesting that Sim identifies as non-mainstream users that are probably not. In summary, and even though this comparison could in turn be considered favorable to Sim (note that previously we assessed against *test nDCG*, not against the *validation nDCG* calculated by Util), the results again support the use of Util to quantify user mainstreamness and mitigate the bias.

## 4.5. DISCUSSION

A key assumption of our approach based on Util is that we can reliably use utility, measured as the accuracy on a validation set, to determine the weight that each user should have in the training process. This implies that the accuracy on the validation set is a good estimate of the accuracy on the test set, which is where the effect will ultimately be assessed. If there was a low correlation between validation and test accuracy, the loss function would apply high weights for users that do not really need it, limiting or even altogether canceling the potential of our approach.

Intuitively, how well validation and test scores correlate is mainly determined by the amount of data. If only a few interactions are involved in the calculation of accuracy, the resulting scores will bear a high degree of noise or random error, thus lowering the correlation. In principle, we would therefore use as much data as possible in the validation and test sets. However, we would generally prefer to use all that data to actually train the model, but we note that the validation scores are somehow part of the training process itself, because they determine the weights.

A balance is therefore necessary, so we need to study the strength of the validation-test scores correlation as a function of the number of interactions in their data partitions. We did this by running the baseline FM model on different data partitions with varying minimum numbers of relevant items in the training set (3, 4, 5 and 10), and validation and test sets (1, 2, 3, 4 and 5 each). The actual split was conducted maintaining proportions (i.e. for the combination of 4/3/3 minimum items per set, a user has 40% of their relevant items for training, 30% for validation, and 30% for testing). We then measured the strength of the validation-test correlation via the RMSE of the scores and their Spearman $\rho$ correlation.

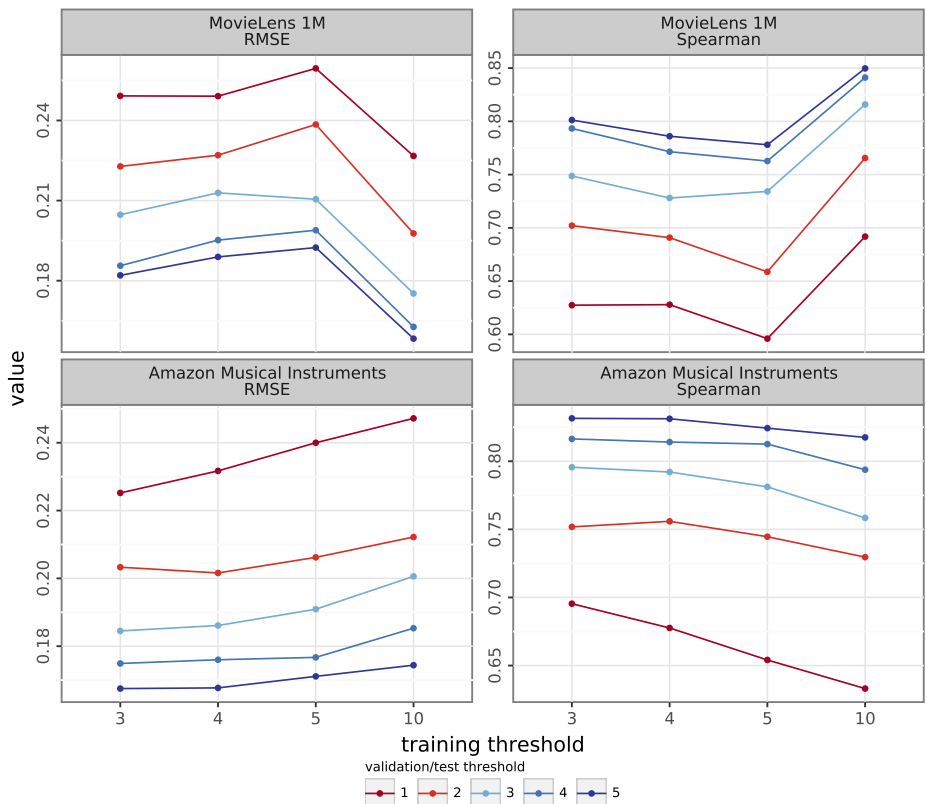

Figure 4.7: Correlation between validation and test scores as a function of the amount of data used for training, validation, and testing, for two sample datasets (most and least dense).

Fig. 4.7 shows that, as expected, the correlation increases (low RMSE, high $\rho$) with the number of relevant interactions used in the validation and test sets. More interestingly, it shows that the amount of training data has a much smaller and varying effect, so despite it being a major factor to maximize model performance, it is not so to robustly assess that performance. The plots indicate that requiring only one or two interactions in the validation set would lead to noisy scores; four interactions seem the bare minimum. As for the training set, the usual practice of having at least as much data as for validation and testing still applies in this context of non-time-aware recommendation.

All in all, our suggestion for this line of research on mainstream bias that works at the individual user level, is to have no less than four items per user in each of the three standard data partitions. Because the strength of the correlation is a key factor in our approach, we decided to require at least five to be on the safe side. In fact, we also ob-served that the effect of cost-sensitive learning in the validation sets is similar to what is reported in Figs. 4.5 and 4.6.

## 4.6. CONCLUSIONS AND FUTURE WORK

In this paper, we tackled the challenge of mainstream bias in CF-based recommendation. The main aspect we focused on is to steer the process of mitigating this bias directly by the utility resulting from the recommendation model and data at hand. For this purpose, we proposed an approach that assigns each user an importance weight during training, with these weights defined in a cost-sensitive manner. By choosing to steer the model directly towards the users that receive low utility, and not towards those that *appear* to be non-mainstream, we avoid the model to focus on users that already receive high utility even if they were not expected to. This way, the model does focus on the niche users that suffer from the bias.

Empirical results show that such models produce a more effective balance of the recommendation utility among the mainstream and non-mainstream users, in a way that is consistent across datasets with varying properties. In addition, we provide suggestions regarding the minimum number of interactions to require when partitioning datasets. Without enough interactions, research on mainstream bias at the level of individual users might produce unreliable results.

For future work, we will first explore other ways to quantify mainstreamness. In the implicit measurement sense, an evident question is whether other metrics such as *AP*, or even the combination of multiple metrics, work better at identifying niche users. Additionally, we can think of ways to make the validation-test correlation robust to issues like sample selection bias, for example via inverse propensity scoring. Another line is to explore more principled approaches for an explicit quantification through an extensive study of the factors that influence mainstreamness, such as the temporal dynamics.

Regarding our cost-sensitive learning approach, we will explore its generality, to see how it works for underlying models other than FM or other ranking frameworks such as pairwise and listwise. We will also investigate the combination of cost-sensitive and adversarial learning strategies to mitigate mainstream bias: cost-sensitive to tell the model where to focus on, and adversary to tell how.

Finally, we note that our focus in this paper has been on the effect of mainstream bias mitigation on the users, but one could wonder about what effect it has on the items. One hypothesis is that non-mainstream users are better served because the less popular items are now more likely to be recommended, so it would be interesting to study whether mitigating one bias amplifies or mitigates other biases, such as popularity or position.

# 5
# REFLECTION AND FUTURE WORK

## 5.1. REFLECTION

We started this thesis by discussing user satisfaction as the prerequisite for the successful deployment and exploitation of recommender systems. While a typical recommender system involves multiple stakeholders, like product vendors and platform operators, whose commercial interests also needs to be met, none of these interests can be met without a solid user base satisfied with the recommendation service and therefore returning to the platform for new transactions. We then also discussed the multi-faceted nature of user satisfaction and concluded that recommendation accuracy is the precondition of user satisfaction. We refer to accuracy as the ability of the system to push the most relevant items for the user to the top of the recommendation list, where the term *relevant* stands for the items that would match the preference profile of the user. Further analysis focused on the consequences of the fact that training a recommender system is essentially a machine learning task, the success of which critically depends on the criterion to optimize for and the presence of imbalance in training data. This analysis led to the definitions of two research questions that steered the research reported in this thesis.

Our **first main research question (RQ1)** was defined as follows: *In order to achieve the best possible accuracy for a broad population of users, should we optimize a recommender system for the criterion we would evaluate it on?* This question addresses an important aspect of the process of learning to recommend, namely that it tries to optimize a given criterion in order to maximize the accuracy across the users. Traditionally, the choice for optimization is aligned with the metric for evaluation, with the hypothesis that such alignment will lead to the model that maximizes the final goal. In Chapter 2, we challenge this hypothesis in the context of ranking-oriented recommender systems with binary user feedback. We look into several popular ranking-oriented evaluation metrics in information retrieval, namely *nDCG*, *AP*, *RR* and *RBP*, and use them directly as the optimization target in two different learning paradigms, namely pairwise and listwise recommendation. For *RBP*, which has been largely overlooked in recommender systems research, we further propose an effective method to optimize for it. Empirical results show that it is indeed not necessary to make the alignment of the optimization and evaluation criteria. Compared to directly optimizing for the metric used for evaluation, it could be more effective to optimize for some other metrics. In addition, the results show that among all four metrics for optimization, optimizing the *RBP*-inspired loss achieves the highest scores regardless of the metric used for evaluation. While the superiority over other optimization alternatives is not always significant, *RBP* still provides a promising alternative metric to optimize for in ranking-based recommender systems. With the results indicating that the advantage in optimizing for *RBP* is achieved for all users, but even more for those who are already well served, *RBP* could be an ideal optimization target in many recommendation use cases focusing on user loyalty. However, if balanced recommendation service quality across the user population is desired, this optimization approach may result in losing the users from the long tail.

A recommender system ideally provides recommendation service optimally tuned to the preference profile of every user. However, the results from Chapter 2 show, once again, that maximizing average accuracy does not guarantee that all users are served sufficiently , even if the best possible optimization criterion is selected. Bias in training data makes it challenging to achieve this balance without extra intervention in the model. Among multiple sources of bias, the *mainstream bias* is an important reason for causing the accuracy imbalance across the user population. This is because the preference information of the users with more data points (e.g., by being more active and sharing preferences with other users) dominates in steering the process of learning the recommendation mechanism. For achieving balanced recommendation quality over the user population, we therefore conducted a dedicated investigation focusing on the imbalance in training data and guided by our **second main research question (RQ2)**: *How to mitigate the mainstream bias in recommender systems, so that different users or user groups share a more balanced recommendation accuracy?*

In Chapter 3 we design the process of learning the recommendation mechanism in a way such that it aims, not only to maximize the recommendation accuracy, but also to ensure a good representation of the intrinsic features of the user profiles independently of the amount of interaction data. The rationale is that, by forcing the model into such improved representation, it creates conditions for the recommendation mechanism to serve underrepresented users better. In order to achieve this, we rely, next to the standard user-item interaction data, on extra information containing intrinsic user features to enhance the user profiling. More specifically, we propose NAECF, which deploys the process of reconstructing user review features as an adversary to collaborative filtering. As such, this model learns both the traditional rating-based recommendation and aims at retaining the intrinsic user features. Empirical results show that the proposed method indeed has the accuracy-balancing effect we look for, that is, it helps non-mainstream users more than the mainstream ones.

While Chapter 3 showed the potential of an adversarial mechanism, which focuses on better user representations, in the mitigation of the mainstream bias. However, asking the model to optimize these representations is not sufficient to achieve the desired balance in recommendation accuracy. In order for the model to focus on the users that should receive better recommendations, we need to point it explicitly to them. To that end, in Chapter 4 we test the idea of cost-sensitive learning from imbalanced data. We now consider recommendation accuracy as an implicit but direct proxy for mainstreamness, and use it to directly intervene by weighing users proportionally to their non-mainstreamness, thus making the cost of failing to serve them well higher in another run of the training process. Next to the fact that the empirical results show the effectiveness of this approach, this investigation also points to its other advantage, namely of being contextualized within the recommendation model being learned. Since the mainstream bias is rooted in training data, it may exhibit itself differently in different recommendation contexts. Therefore, instead of proposing a model-agnostic method relying on some generic hypotheses on mainstreamness, we make a case in this chapter for coping with mainstreamness dependent on the data and model at hand.

## 5.2. FUTURE WORK

Based on the insights provided in this thesis, we now point out some ideas that can serve as a basis for future research in the field of recommender systems. We describe these ideas in the following, according to the main topics addressed in this thesis.

68 5. Reflection and Future Work### 5.2.1. ON THE CHOICE OF OPTIMIZATION TARGETS

In Chapter 2, we show empirically that optimizing for *RBP* is more effective than the usual alternatives in maximizing the recommendation accuracy. Full understanding of this effectiveness could, however, be helped by a theoretical analysis of the superiority of *RBP*. Such analysis would not only further strengthen that choice as a strong optimization target, but also steer the discussion and future development of new optimization targets and evaluation metrics.

In addition, we note that the research presented in Chapter 2 is limited to the case of binary relevance, motivated by the vast amount of implicit feedback in recommender systems indicating the presence of an event such as a click or playback. However, multi-level relevance, for instance via graded ratings, is still crucial in recommender systems and provides more fine-grained data regarding user preferences. While training processes for binary relevance are simpler, turning graded relevance into binary incurs a loss of information. Moreover, the dichotomization of relevance may contribute noise to the training process if items of low utility are treated as relevant [133]. Previous research has already explored the optimization of Graded Average Precision [115] and Expected Reciprocal Rank [116], but choosing the optimal target for training in the multi-level relevance scenario is still an open question that may steer follow-up work.

### 5.2.2. ON OPTIMIZING FOR MULTIPLE TARGETS

In Chapter 2, we set our aim at investigating the effectiveness of optimizing for one single target, which is an accuracy-oriented evaluation metric derived from information retrieval. But even when a single metric like *RBP* appears to maximize accuracy, it is still possible that optimizing for multiple metrics simultaneously achieves even better performance either on average or across the user base, as different metrics consider different aspects of the recommendation list and they could therefore complement each other.

While the optimization of accuracy was motivated by the hypothesis that it is a prerequisite for user satisfaction, beyond-accuracy criteria such as novelty, diversity, serendipity and self-actualization remain important as well. This evidence the necessity to see even the optimization of user satisfaction only as a multi-objective optimization problem in which different objectives may conflict with each other. With the power of advanced machine learning methods such as adversarial learning and multi-arm bandits, this line of research could be further extended to the search for a good balance between accuracy and beyond-accuracy criteria across different stakeholders. Given that the ultimate goal of recommender systems is to satisfy the interests of everyone involved, it would be worthwhile to investigate how, for example, maximizing the recommendation accuracy for the user can be combined with maximizing the diversity for the product vendor (e.g., in order to make sure the whole product stock is offered on the market). In addition, Chapter 3 made it evident that we also need to consider other factors like bias in data, which are related to balancing the quality of the recommendations across users. One of the interesting research directions would therefore be to investigate further how to make this multi-objective optimization even more effective.

### 5.2.3. ON UNDERSTANDING MAINSTREAMNESS

In chapters 3 and 4, we used the recommendation accuracy obtained per user in a given context as a proxy of mainstreamness, but we did not investigate what causes the deficient accuracy. The reason can lie in the deviating patterns of interacting with the items in the past compared to the majority of the users, or in the deviating taste focusing more on the long-tail items, or in the imbalances in data availability due to differences in activity, or in something else. Understanding the reasons for deficient recommendation service would lead to a better understanding of mainstream bias, which would ultimately help fine-tune the mitigation effort to the varying characteristics of non-mainstream users. Furthermore, if we look at the dynamic recommendation scenario in which users change their interaction behavior and taste over time, the distribution of mainstreamness also varies, which needs to be taken into account by updating the recommendation mechanism over time.

### 5.2.4. ON MITIGATING THE MAINSTREAM BIAS

In chapters 3 and 4, we contributed to the research on mitigating the mainstream bias from two different perspectives: one method focusing on the identification of the users *who* should receive better recommendations, and the other one mainly on *how* to help the model produce those better recommendations. The next intuitive step would be combining both in a unified framework to benefit from the expected synergy. Second and more importantly, our contributions focus more on recommendation models, which is the in-processing stage in getting recommendations. However, this is not necessarily the only choice. With bias being rooted in data and models incorporating the interventions to mitigate this bias, it could be productive to manipulate the data prior to training, resulting in less imbalanced data. With the emergence of fairness-aware [29] and self-supervised data augmentation [143], designing such data manipulations to compensate the original imbalance would offer more freedom than designing model architectures. For the follow-up research, a possible direction is to first see whether, and how much data augmentation could help the underrepresented users, for which it would be useful to now what the source of under-representation is, as pointed out in the previous section. If it comes from users being less active, then synthetic data could help enhance the existing user-item interaction data. However, if the reason lies in taste difference, even a full user-item matrix would bias the model, because for a small group of users with deviating tastes insufficient neighboring users could be found for reliable collaborative preference modeling. It is therefore worth investigating when to do data pre-processing and when to do model-based in-processing for mainstream bias mitigation, or even when and how to combine both effectively.

# LIST OF FIGURES

# LIST OF TABLES

# SUMMARY

Recommender Systems have drawn extensive attention in recent decades, because they are a powerful tool with the potential to help several business stakeholders –including end users, sellers, and platform providers– through personalized recommendations. The most important factor to make a recommender succeed is user satisfaction, which is largely reflected by the recommendation accuracy. Therefore, one primary question in recommender systems research is how to make all users enjoy good recommendation accuracy. This thesis dives into this question from two different perspectives that, unfortunately, are at tension with each other: achieving the maximum overall recommendation accuracy, and balancing that accuracy among all users.

The first part of this thesis focuses on the first perspective, that is, maximizing the overall recommendation accuracy. This accuracy is usually evaluated with some user-oriented metric tailored to the recommendation scenario, but because recommendation is usually treated as a machine learning problem, recommendation models are trained to maximize some other generic criteria that does not necessarily align with the criteria ultimately captured by the user-oriented evaluation metric. Recent research aims at bridging this gap between training and evaluation via direct ranking optimization, but still assumes that the metric used for evaluation should also be the metric used for training. We challenge this assumption, mainly because some metrics are more informative than others. Indeed, we show that models trained via the optimization of a loss inspired by Rank-Biased Precision (RBP) tend to yield higher accuracy, even when accuracy is measured with metrics other than RBP. However, the superiority of this RBP-inspired loss stems from further benefiting users who are already well-served, rather than helping those who are not.

This observation inspires the second part of this thesis, where our focus turns to helping non-mainstream users. These are users who are difficult to recommend to either because there is not enough data to model them, or because they have niche taste and thus few similar users to look at when recommending in a collaborative way. These differences in mainstreamness introduce a bias reflected in an accuracy gap between users or user groups, which we try to narrow.

Our first effort consists in using side data, beyond the user-item interaction matrix, so that users and items are better represented in the recommendation model. This will be of benefit specially for the non-mainstream users, for which the user-item matrix alone is ineffective. We propose Neural AutoEncoder Collaborative Filtering (NAECF), an adversarial learning architecture that, in addition to maximizing the recommendation accuracy, leverages side data to preserve the intrinsic properties of users and items. We experiment with review texts as side data, and show that NAECF leads to better recommendations specially for non-mainstream users, while at the same time there is a marginal loss for the mainstream ones.

Our second effort consists in explicitly signaling to the training process what users it should focus on, that is, the non-mainstream ones. In particular, we propose a mechanism based on cost-sensitive learning that weighs users according to their mainstreamness, so that they get more attention during training. Here we argue for not quantifying mainstreamness directly, but rather its effect, and therefore weigh users depending on how well they are served by a vanilla recommendation model. The result is a recommendation model tailored to non-mainstream users, that narrows the accuracy gap, and again at negligible cost to the mainstream users.

**5**

# ACKNOWLEDGEMENTS

Back in the early morning of September 3, 2017, a 10-hour flight took me from Beijing to Schiphol. Taking off during the night, my sleepiness made me a bit blunt. I did not really feel much difference until got my mobile phone connected to the WiFi in the airport. When the web page was automatically redirected to Google and I did not see the familiar error 502 popping up, I finally realized the important fact: I am in the Netherlands. The life afterward was largely dedicated to giving birth to this booklet. Like all the other birth processes, it is a mix of pain and joy, with the help of quite a few people. At the last mile before entering into my D-Day, it is time to say thanks mainly in English, after doing so in modern Chinese for the Bachelor's thesis and in classical Chinese for the Master's thesis.

Let me first start with my PhD committee. Alan, you stay at the top of the list. It is also a good coincidence that I am now writing in Erlangen, the city you stayed for four years as a student. Up until now I still remember the afternoon of August 22, 2016 in Beijing. I almost finished my daily work at Toshiba Medical Systems. Before stepping out of the office to go back home, I gathered all my courage to send you the cold email requesting information about a potential PhD position at the Multimedia Computing Group (MMC). To be honest, that was just my second try on this journey, so I did not really have a high hope for a positive reply, even though I know TU Delft is my first choice. Surprisingly, 5 minutes later, when I was waiting for the bus to go back home, I got a positive reply from you. Maybe for you, that was just a normal custom, but you might never know how encouraging it was for a young guy who did 3 jobs at the same time while doing the PhD application in the summer of a crowded city. After coming to Delft, I know I get the best promotor I can have: a rigorous and strong true researcher, a role model in dealing with almost all issues correctly, and a supportive, nice, and patient person to work with and learn from. In my worst times, you were the person to tell me to take a break and calm down; when I was depressed, you told me to relax and just do something not related to research; though when the real research was done, you were also the person to push for better quality. Thanks Alan, meeting you in Delft is one of the luckiest things I can imagine in my life. Hvala.

Julián, you are the hero always behind me, from all perspectives. Our discussion ranges from football news to research, and then even to some personal communications. I am always impressed by the perfect combination of your scientific rigor and willingness to support rather than criticize. In the early years when I kept messing things up, you were still there to help, from data management to analysis, and then to research methodology. Do you still remember? After the submission to SIGIR'20, even I apologized to you for making you work on such a terrible draft, but you just told me "Why do you apologize? This is my job." That was one of the most moving moments during my PhD. I will also never forget those nights before the deadlines you stayed together with me, and will remember all the things I learned from you, both as a great person and a great supervisor. Your way of work shaped me, and the current skill set I have for daily work is almost fully from you. For now, I always aim at doing one-third of what you did to me to my current colleagues, though even such a proportion is a challenge for me. After the acceptance of our paper to SIGIR'21, Alan told me to build a statue for you, and I just wanna let you know that it is always in my heart. I am also grateful you agreed to become the witness of my wedding and told me you hoped to leave some positive memories of you in my heart apart from the PhD grind. As I told you then, there are ups and downs, but never any negative memories about you. With our occasional discussion over football, what I wanna say is not only Gracias, but also Hala Madrid.

To all other independent committee members, Prof. dr. ir. Geert-Jan Houben, Prof. dr. Martha Larson, Prof. dr. Markus Schedl, Dr. Enrique Amigó, Dr. Sole Pera, and Prof. dr. ir. Alessandro Bozzon, thanks for reviewing this booklet and providing constructive feedback to make it better. It is my honor to have such a diverse and admirable committee. Especially, Martha, you did not directly get involved in my research, but the suggestions I got from you in Delft, Vancouver, and Amsterdam always benefit me, and I am so proud that one of my technical chapters follows your path and legacy in Delft. Markus, you are one of the earliest pioneers in the area of mainstream bias in recommendation. I tried to talk to you even since RecSys'19 in Copenhagen, but never dared to really do this. Now it is finally to get the chance. Sole, our correspondence dates back to early 2017 when you interviewed me together with Michael in Boise. Though I did not really accept the offer from PIReT, it is amazing to finally meet you in Delft. It is a beautiful circle to get interviewed by you before the start of my PhD and then have you in my defense to conclude it.

MMC is my home in Delft. What I have here is not only a research base, but also excellent minds to discuss with. Cynthia, you also interviewed me before I came to Delft. These years, in addition to absorbing your idea of caring about the evolution of user interest, I also wanna express my gratitude for your attention to the life situation of PhD candidates, especially the help in accommodating me at RecSys'21 when I was quite hard up. Huijuan, on my first day in the office you asked me the scientific question of my PhD. You are almost the first person to remind me of shaping the idea from the helicopter view before thinking about details. As the "parent" of Chinese PhD candidates in the group, you always have the magic to shed light on my life after talking to you. Elvin, we do not talk too much about research, but I am deeply impressed by the way you teach, as well as telling me about the negative facts of the PhD job market in the Netherlands. Luckily and unluckily, it turned out to be true, but I got better prepared because of you; Odette, I am always impressed by how you organize events at different levels and make all of them function perfectly; Jorge, the talk with you happened mostly during lunch. I should have put the sentence heard from you "All models are right, but some of them are useful" into my propositions if there was no plagiarism check; Zhengjun and Pablo, thanks for introducing your amazing work to me to broaden my horizons. Special thanks to Saskia: you did everything regarding administration perfectly, and managed to resolve all potential concerns, sometimes even before I realized them. Thank you so much and I never take it for granted.

Now turning to my amazing office mates. Manel, what a person you are! Started our PhDs on the same day, submitted papers together to RecSys'18, SIGIR'19, RecSys'19, and WSDM'21, and traveled together to Vancouver, Gothenburg, Copenhagen, and Am-sterdam, we shared so many memories. Furthermore, we were even colleagues twice outside the university, as student volunteers of RecSys together. We felt each other and encouraged each other throughout the journey, witnessed the ups and downs of each other, and become good friends. As the manager of my wedding ceremony and one paranymph of my defense, you further make our friendship increasingly solid. Xiu-xiu, my Chinese elder in the group, you impressed me with your faith in research and highly disciplined lifestyle. Those talks in the office, on the way back home, and in the sports center are so precious, even though you are never a fan of my hometown, where you also stayed for many years. Matteo, another guy in Julián's army, with a typical Italian lifestyle but is almost from Slovenia. You might be more like Julián than me in the stubbornness of diving deep into one topic. Thanks for being another paranymph. Alberto and Omar, we share quite some common topics in chitchatting, from language acquisition to sports, and then to geography. It is always good to have peers listening to each other like you. Happy for Omar, as your home folk Fernando Alonso performs so well at the age of 42; sorry for Alberto, as Pogba did not play much after returning to Juve before almost getting the red light of his career. I do not want to build my happiness over your pain, but it is indeed a relief for me that this did not happen two years ago. Jaehun, the Encyclopedia of Research, you helped me so many times in both research and subtle implementation details. You are also the person to make me love bibimbap. Tingting, another connection between MMC and Taiyuan, the talented generative AI expert, your work was ahead of time. I am also amazed by how well-organized your life and code can be, and how authentic your homemade cuisines are. It is a pity that you met a very premature Roger as the research collaborator. I owe you a co-authored paper. Other Chinese peers, Cheng'en, Cunquan, Li, Lingbo, Maosheng, Siyuan, Shilun, Tianrui, Tianqi, Tianyi, and Yuanyuan, I enjoyed talking to you, playing badminton with you, and more importantly, having delicious foods together at different places. Soude, Raynor, Ernestasia, Karthik, Babak, Sandy, Andrew, Bishwadeep, Mohammed, Patrick, and Antony, all the talks with you are either informative, interesting, or both. I regret we could not have it more.

My neighbors in Delft, Alice, Kailun, Xiujie, and Yancong: I cannot remember how many times we get together for hotpots, delicious snacks (thank you Alice), sports, and board games. We talked about almost everything, celebrated the Chinese New Year together, and had a lot of interesting online conversations even if we were just meters from each other. You made me feel at home all the time, and I cannot describe such a special friendship only via words. However, to Yancong: we were neighbors for over 8 years and became colleagues in two different institutes. You are not only a great friend, but also the best chauffeur and wedding assistant when needed. Getting to know a guy like you needs quite some luck, and I know I am lucky.

Zhejiang University alumni in NL, you gave me a lot that I never expected to have before going overseas. Though a bit peculiar, I enjoy talking to many of you using the words and notions that could be correctly understood by us. Beien and Brother Bricks (Xuan), it is always easy to sit with you and find something interesting to talk about. I am also happy to poach Beien's skill in cooking his signature Zhajiangmian. Thanks man, I become more Beijing-like now on top of your endorsement of my Beijing accent, while yourself is more away from your Beijing aboriginal identity. Xuan, long time without a badminton game with you. Next time let us see whether you get closer to Taufik Hidayat. Well, I mean badminton skills, not the weight. Fandy, thanks for teaching me, Beien, and Xuan something we are too old to understand or master. Zhenyu and Qu, you asked me questions about computer science before, but now with your extensive experience in industry, I might need your help. Other friends in this community, Yupeng, Litian, Jingyi, Meng, Zhe Hou, Xuan, Tianyuan, Haoyuan, Xiaowei, Longxiang, Fangzhou, etc, thank you all for the time spent together. It is always awesome to have such a large group of people with the same connection built from over 8,000 kilometers away.

My adidas family, maybe you are not directly related to my PhD, but the experience working with you makes me a better data practitioner, business reporter, and possibly also team leader. Alan, Steffen, and Mahendra, thanks so much for providing this great position and making it available in the Netherlands. You also gave me the freedom to do what I am keen on, which is quite rare in the business context. Jozsef, I like all the technical discussions between us. Our working relationship is never like the traditional product owner vs. data scientist mode. By observing how you code and present, I even became a better researcher after going into the industry. Mohit, I would never expect a better engineer than you to get all things done so well. Varun, Narender, Arpit, Manish, Shaghil, Ramendra, Krishan, Disha, Abhi, Anis, Ali, Nathalie, Katharina, Murtuza, Rahul, Philipp, Lucas, Jorge, Arne, you all helped me explicitly or implicitly in making me transfer successfully from academia. No matter how senior you are or where your expertise lies, you are all great people I truly want to work with. Your support means more than you think to help me finish my PhD in a multi-tasking mode for over one year. Other peers in the Amsterdam office, Rahel, Kshitij, Sreeram, Garima, Jeroen, Jean, Sharon, Shuxin, Artëm, Abhinav, Madhavi, Uday, Federico, João, Yongrui, Nikhil, Ishman, Hina, Arijeet, Prerana, Abhigyan, Aisha, and other names I may not mention here, we work hard, play (table tennis) harder, and party hardest. You make me have the true feeling of society. Thank you all for being with me for part of the PhD journey.

Before going more personal, another party to thank: Manchester United. To be honest, at least during these years, I do not think it should be I thank you. You should thank me for my everlasting support even if you did not achieve anything important since I am overseas! The Last Europa League title was 6 years ago, the last Premier League title was 10 years ago, and the last Champions League title was 15 years ago! This is not good, but you are still an important part of my personal life outside of research and daily work, and I got the chance during my PhD to visit Old Trafford, the Theatre of Dreams, 3 times. It is a dream come true for me, and the love since 1998 is hard to vanish. An era has a beginning, and will also be with an end. This low era will finally be gone, and I hope you can give me more joy after my PhD, just like what it was before my Bachelor's. It is time for me to embrace a new chapter of life, and maybe also time for you to get the glories back.

Now family time. A PhD means several years of career for many people, and for me the endeavor of 25 years since I was a pupil, but for you, Mom and Dad, I know it is a process of 31 years. Your unconditional love and support is the thing I always have, no matter where I am and what I do. Without your education, I may never have the courage to cross half of China for a Bachelor's, and then further be a continent away from you. The most important thing I get from you is to always keep rational and go for the long-term goals, even if it means a loss for now. The loss is already there: as a guy doing PhD during the COVID, I have not seen you for 4 years. It is part of the price for a PhD, but at least it pays off. Also, thank my cousin for hosting me several times on my visit to Paris.

Finally you, Yang. I am so happy to call you my wife in this paragraph. Not so many people can stand a long-distance relationship for over 5 years with more than 1 year under the COVID travel ban, but we made it. I promise I will repay your support, trust, and love with the rest of my life. I will not swear more, but just allow me to repeat what I said to you on August 28, 2023, the day we got married:

*I give you this ring as a symbol of my willing commitment to respect, cherish, support, love, and always remain faithful to you.*

<div style="text-align: right;">
Roger Zhe Li<br>
October 2023<br>
Erlangen and Herzogenaurach, Germany
</div>

# Curriculum Vitæ

Roger Zhe Li was born on May 3, 1992 in Taiyuan, People's Republic of China. From 2010 to 2014 he pursued a Bachelor of Engineering in Information and Communication Engineering in Zhejiang University, Hangzhou, People's Republic of China.

Right after graduation, he stayed in the same country but went to Tianjin University to keep majoring in Information and Communication Engineering as a Master's student, with research focus on computer-aided diagnosis on breast cancer and machine learning on imbalanced data. He got his Master of Engineering degree in 2017 with *cum laude*. During this period, he also had several internship experiences. In 2016, he worked as an intern at Toshiba (now Canon) Medical Systems China on medical image algorithms, and also at the National Laboratory of Pattern Recognition, Institute of Automation, China Academy of Sciences on data mining of electronic medical records. In 2017 he worked as an intern at SenseTime on the R&D of social recommender systems.

Since September 2017 he pursues his PhD at the Multimedia Computing Group at Delft University of Technology, supervised by Prof. dr. Alan Hanjalic and Dr. Julián Urbano. His PhD topic covers learning to rank and user fairness in recommender systems, and his work got published on prestigious venues including ACM RecSys, WSDM, SIGIR, and ICTIR. He actively provides academic service, and served as a PC member of ICTIR'23, an external reviewer of SIGIR'23, TheWebConf'23 and SIGIR'22, and a student volunteer of ACM RecSys in 2018 and 2021. He also reviews for journals including IEEE Systems Journal and Neurocomputing.

In January 2022, Roger started working as a Senior Product Owner at adidas, based in Amsterdam, the Netherlands. There he leads the development of the internal computer vision-based item feature store. His work spans self-supervised learning, image style transfer and generation, image retrieval, and cloud-based machine learning engineering.

# LIST OF PUBLICATIONS

## PUBLICATIONS AS THE (*de facto*) FIRST AUTHOR

12. **Roger Zhe Li**, Julián Urbano, and Alan Hanjalic. *Mitigating Mainstream Bias in Recommendation via Cost-sensitive Learning.* In Proceedings of the 9th ACM SIGIR / 13th International Conference on Theory of Information Retrieval (ICTIR'23). Association for Computing Machinery, New York, NY, USA, 135-142.

11. **Roger Zhe Li**, Julián Urbano, Alan Hanjalic, 2021. *New Insights into Metric Optimization for Ranking-based Recommendation.* In Proceedings of the 44th International ACM SIGIR Conference on Research and Development in Information Retrieval (SIGIR '21). Association for Computing Machinery, New York, NY, USA, 932–941.

10. **Roger Zhe Li**, Julián Urbano, Alan Hanjalic, 2021. *Leave No User Behind: Towards Improving the Utility of Recommender Systems for Non-mainstream Users.* In Proceedings of the 14th ACM International Conference on Web Search and Data Mining (WSDM '21). Association for Computing Machinery, New York, NY, USA, 103–111.

9. **Zhe Li**. 2018. *Towards the next generation of multi-criteria recommender systems.* In Proceedings of the 12th ACM Conference on Recommender Systems (RecSys '18). Association for Computing Machinery, New York, NY, USA, 553–557.

8. Wei Lu, **Zhe Li**, and Jinghui Chu. 2017. *Adaptive Ensemble Undersampling-Boost.* Journal of Systems and Software. 132, C (October 2017), 272–282.

7. Wei Lu, **Zhe Li**, Jinghui Chu. 2017. *A Novel Computer-aided Diagnosis System for Breast MRI based on Feature Selection and Ensemble Learning,* Computers in Biology and Medicine, 2017, 83: 157-165.

6. **Zhe Li**, Wei Lu, Hang Min, Jinghui Chu, 2016. *Application of machine learning algorithms in breast tumor detection,* Computer Engineering & Science, 2016, 11: 2303-2309.

## PUBLICATIONS AS CO-AUTHOR

5. Libao Yang, **Zhe Li**, Guan Luo, 2016. *MH-ARM: a Multi-mode and High-value Association Rule Mining Technique for Healthcare Data Analysis,* International Conference on Computational Science and Computational Intelligence (CSCI), 2016, 122-127.

4. Jinghui Chu, Zerui Wu, Wei Lu, **Zhe Li**, 2018. *Breast Cancer Diagnosis System Based on Transfer Learning and Deep Convolutional Neural Networks,* Laser & Optoelectronics Progress, 2018, 55(8): 081001.

3. Wei Lu, Weixian Deng, Jinghui Chu, **Zhe Li**, 2018. *Arrhythmia Classification Based on Feature Selection Method of S-transform,* Journal of Data Acquisition and Processing, 2018, 33(2): 306-316.

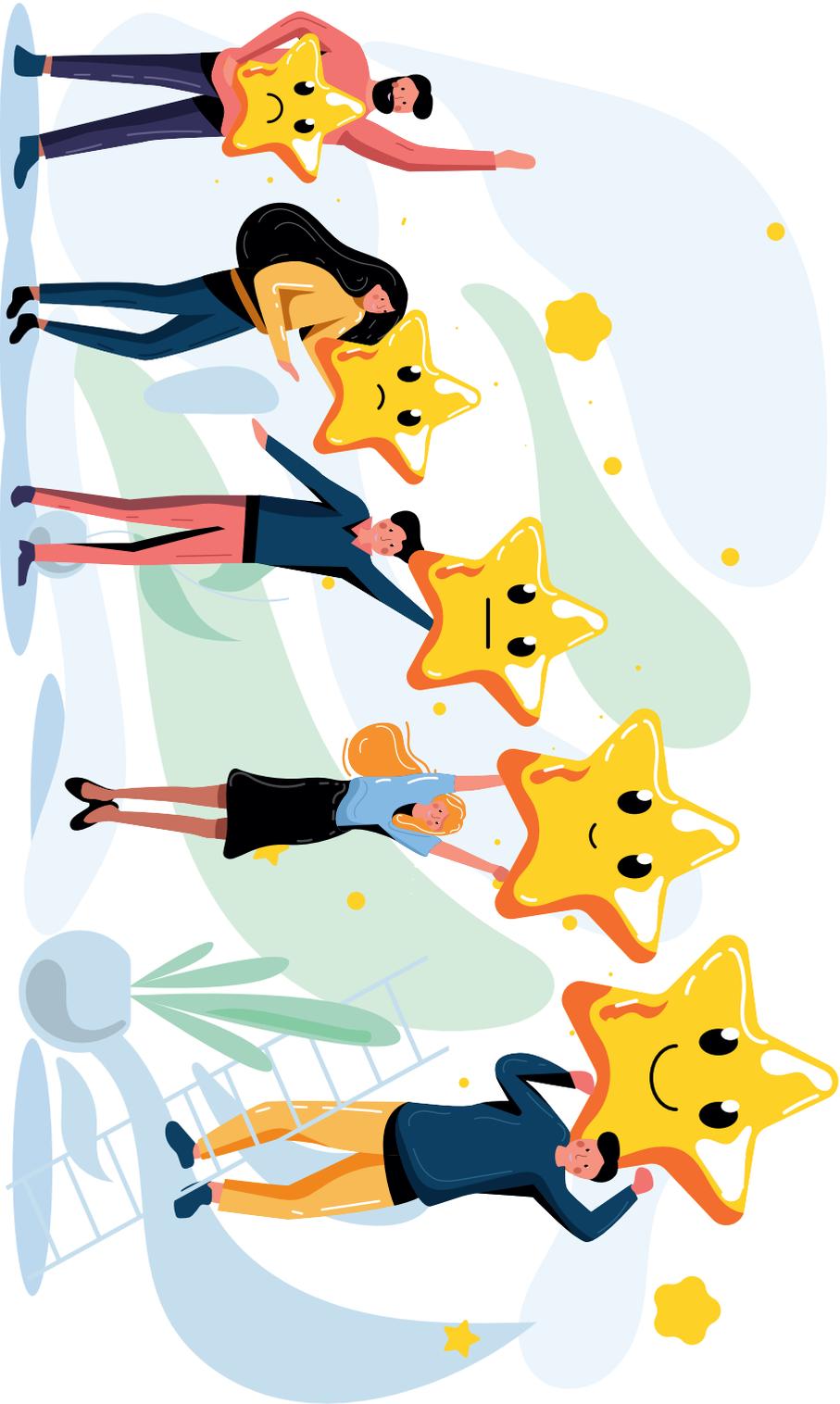